\documentclass[a4paper,fleqn]{cas-sc}

\usepackage[square,numbers,comma]{natbib}
\usepackage{mathtools}
\usepackage{subcaption}
\usepackage{calligra}
\usepackage{ulem}
\usepackage{algorithm}
\usepackage{algpseudocode}
\usepackage{siunitx}
\usepackage{cleveref}
\usepackage{tabularx}
\usepackage{gensymb}
\usepackage{algorithm}
\usepackage{algpseudocode}
\usepackage{hyperref}

\def\tsc#1{\csdef{#1}{\textsc{\lowercase{#1}}\xspace}}
\tsc{WGM}
\tsc{QE}
\tsc{EP}
\tsc{PMS}
\tsc{BEC}
\tsc{DE}
\begin{document}
\let\WriteBookmarks\relax
\def\floatpagepagefraction{1}
\def\textpagefraction{.001}

\shorttitle{Evolutionary Cold Plate Design}

\shortauthors{Ignuta-Ciuncanu et al.}

\title [mode = title]{Hotspot-Aware Evolutionary Cold Plate for Pouch-Cell Thermal Management: Design and Experimental Validation}                      
\author[1,2]{Matei C. Ignuta-Ciuncanu}[type=editor, auid=000, bioid=1]
\cormark[1]
\ead{mi3618@ic.ac.uk}
\credit{Conceptualization, Methodology, Software, Analysis, Writing - Original Draft, Visualization}
\affiliation[1]{organization={Department of Nuclear Engineering, University of California, Berkeley, CA},
    % addressline={Exhibition Rd}, 
    % city={London},
    % citysep={}, % Uncomment if no comma needed between city and postcode
    country={USA}}

\affiliation[2]{organization={Sustainable Energy Technology and Turbomachinery Lab, Imperial College London},
    city={London},
    postcode={SW7 2AZ}, 
    country={UK}}

\affiliation[3]{organization={Electrochemical Science and Engineering Group, Imperial College London, SW7 2AZ},
    % addressline={Exhibition Rd}, 
    % city={Cambridge, MA},
    % citysep={}, % Uncomment if no comma needed between city and postcode
    country={UK}}

\author[2]{Chris Noon}[type=editor, auid=000, bioid=1]

\credit{Conceptualization, Methodology, Design}

\author[2]{Ricardo F. Martinez-Botas}[%
   % role=Supervisor,
   ]
\credit{Conceptualization, Supervision, Writing - Review \& editing}
\cortext[cor1]{Corresponding author}

\author[3]{Monica Marinescu}
\credit{Conceptualization, Methodology, Writing - Review \& editing}

\begin{abstract}
A cold plate for a lithium-ion pouch cell must distribute cooling across a nonuniform heat source without imposing excessive hydraulic losses. This work combines an evolutionary constructal design procedure with a pseudo-three-dimensional model that couples spatially resolved battery heat generation to cold-plate flow and heat transfer. An organic network selected from the simulated design set was machined in aluminium and assembled with a commercial pouch cell. A separate numerical benchmark compares that network with its straight-channel reconstruction and two parametric plates under matched thermal and flow conditions. At the selected simulated condition, the organic network requires approximately half the simulated pumping power of the fine parametric plate while reducing the battery-surface temperature-nonuniformity objective. The reconstruction retains 60\% of the thermal-hydraulic benefit, while reducing manufacturing time by an order of magnitude. Experiments on the machined organic plate measure cell temperatures, coolant temperatures, and pumping power across four coolant-flow settings and three nominal inlet-temperature conditions. The resulting operating-condition map distinguishes reductions in mean and maximum cell-temperature rise from the power required to circulate coolant and identifies condition-dependent trade-offs. Together, the numerical benchmark and experimental map establish a design-to-hardware workflow and characterize the thermal--power choices available for pouch cells. The results of this workflow are readily extensible for any form factors.
\end{abstract}


\maketitle
\section{Introduction}
Lithium-ion pouch cells are widely used in applications such as electric vehicles (EVs), portable electronics, and grid-scale energy storage due to their high energy density, lightweight design, and flexibility in form factor. 
Excessive non-uniform heat generation during charge and discharge cycles can lead to accelerated degradation \cite{Ardani2018}, capacity fade \cite{Ning2003}, and even safety hazards such as thermal runaway \cite{Wu2015}. 
Moreover, battery lifetime extension through optimized thermal management has been shown to significantly reduce both life cycle costs and carbon footprint, making EV adoption more economically and environmentally viable \cite{Lander2021}.
Therefore, maintaining a uniform and controlled temperature across the battery surface is crucial for ensuring long-term reliability and performance \cite{Bandhauer2011}. 
Various cooling strategies have been explored, including air cooling \cite{Taeyoung2018}, phase change materials \cite{Gungor2024}, and liquid cooling systems \cite{Gungor2022}. 
Among these, liquid cooling via cold plates has emerged as one of the most effective methods due to its superior heat dissipation capability and ability to maintain low and uniform temperatures under high-C discharge \cite{DONDELEWSKI2020}.
In fact, optimizing thermal management has been shown to reduce life cycle costs by 27\% and carbon footprint by 25\%, further reinforcing the advantages of surface cooling over air cooling in maintaining stable operating conditions \cite{Lander2021}.

While liquid cooling is an effective thermal management strategy, conventional cooling designs often overlook the complex interactions between thermal processes in the cold plate and electrochemical processes within the battery. 
Many computational approaches, particularly those using topology optimization (TO), rely on simplified 2D models that fail to capture the non-uniform heat generation resulting from the coupling of electrochemical and conjugate heat transfer.
The assumption of uniform heat generation \cite{Mo2021} and the fluid-domain based decoupled approach \cite{he2023a} can lead to inaccuracies in predicting temperature distribution \cite{He2024} and performance \cite{Alexandersen2020}. 
Furthermore, most studies neglect the bidirectional coupling between thermal and electrochemical behaviour, where local temperature variations influence electrochemical reactions, and in turn, electrochemical activity affects heat generation \cite{Wu2015}. 
The issue is compounded by the lack of detailed mapping of the battery and cold plate surface temperatures in existing publications \cite{Sun2023, Ji2023, Wu2024}, which limits the understanding of how thermal management strategies influence the behaviour of the entire system.

As a gradient-based method, topology optimization (TO) can lead to local optima solutions, limiting its ability to explore the full design space \cite{Bejan2023b}.
In contrast, constructal design principles focus on optimizing flow paths and heat dissipation through natural dendritic geometries, offering a more effective approach to thermal management \cite{Bejan1997, Bejan2002}. 
Unlike traditional design approaches that rely on fixed \textit{end-design} objectives, constructal design embraces evolution, enabling systems to adapt and morph continuously, as opposed to going in the predefined direction of gradient-descent \cite{Bejan2023a, Bejan2023b}.
This flexibility allows for the creation of more effective, adaptive flow architectures that better align with the dynamic and complex nature of thermal management in energy storage systems.

Using machine learning algorithms, generative design allows for the exploration of a broader range of configurations to create more efficient and adaptable solutions. 
Recent advances in computational techniques have improved the ability to integrate modeling, optimization, prototyping, and testing within high-throughput workflows, making them especially effective for processing multi-modal and hierarchical information \cite{Fromer2022, Shen2022, Raabe2023, Khatamsaz2023}.
These approaches, such as the use of generative adversarial networks (GANs) and graph-based representations, have been applied to optimize material properties and mimic natural growth processes to create intricate microstructures with targeted properties \cite{Shen2022}. 
By applying constructal and generative design principles, our group developed non-uniform configurations that enhance heat conduction and streamline the design of dendritic flow networks \cite{IgnutaBotas2025}. 
This methodology allows systems to evolve and adapt to dynamic performance needs, improving their freedom-to-morph.

Thermal-electrochemical bidirectional coupling is crucial because temperature variations directly impact reaction kinetics, ionic conductivity, and degradation rates, while electrochemical activity simultaneously influences heat generation, creating a complex feedback loop that must be accounted for in accurate thermal management strategies \cite{Wu2015}. 
Despite this, most existing thermal management studies rely on simplified models rooted in either thermal design or electrochemical literature. 
Thermal design studies typically treat the battery as a uniform heat source and approximate the cold plate as a fluid domain with effective-medium properties, failing to account for the response of the battery to the ambient. 
Conversely, electrochemical models impose the cold plate as a uniform boundary condition, neglecting its role in shaping local temperature variations and, consequently, electrochemical performance.
While multi-physics 3D simulations can capture these interactions with high fidelity, they are computationally expensive and impractical for optimization due to the high resolution required across multiple scales. 
To address this, a pseudo-3D approach provides a balanced solution by accurately resolving key thermal-fluid interactions while simplifying less critical dimensions, significantly improving computational efficiency without sacrificing essential accuracy.

Thermal management strategies often ignore finite-time and finite-size thermodynamics, which are crucial under real operating constraints \cite{Bejan1996}.
Most models assume reversible, quasi-static behaviour. 
They miss irreversibilities like entropy generation from finite temperature gradients and viscous losses in flow paths \cite{bejan_entropy_1982}. 
In cold plates, increasing heat removal comes at the cost of increased pressure drop. 
This highlights a key trade-off: better cooling versus higher fluidic losses, well-known in the heat exchanger literature \cite{Bejan1978}. 
To manage this, thermal optimization must treat the battery and cold plate as a single thermodynamic system, and minimize total entropy generation.

Moreover, most optimization studies rely solely on simulations without experimentally validating the temperature distribution on the battery surface, often limiting measurements to inlet and outlet fluid temperatures.
As a result, critical real-world factors, such as imperfect contact resistance and other operational variations, are overlooked, leading to potential mismatches between predicted and actual performance.
This gap restricts the understanding of localized heating effects and hinders the accurate assessment of thermal gradients that influence electrochemical performance.
By directly mapping surface temperatures, our approach provides a crucial benchmark for verifying simulation accuracy, ensuring that thermal designs meet the requirements of electrochemical systems.

This study makes three key contributions to advancing the design of energy systems with lithium-ion batteries. 
First, we apply constructal design principles and evolutionary algorithms to optimize cold plate designs, balancing thermal and fluid losses. 
Second, we incorporate a coupled thermal-electrochemical model that captures the bidirectional interactions between conjugate heat transfer in the cold plate and the heat generation and electrochemical processes within the battery. 
Finally, we experimentally validate the optimized designs by mapping the surface temperature distributions across the battery and cold plate interfaces. 
By comparing experimental and predicted results, we show that our approach leads to more reliable and effective thermal management strategies, providing a clear direction for the thermal-electrochemical co-design of energy storage systems.

\section{Methods}
\subsection{Evolutionary Constructal Design Approach}
In this work, we build upon previous efforts by employing generative constructal design, which seeks to optimize flow by emulating natural, hierarchical geometries. 
This methodology, summarized below, leverages machine learning techniques for generating and evolving diverse flow networks; for details regarding the implementation, the reader is referred to \cite{IgnutaBotas2025}.
To address the area-to-point conduction problem \cite{Bejan1997} and efficiently distribute flow from a finite area to a point, we employed an evolutionary process using both supervised and unsupervised models to generate diverse flow network configurations.
A supervised model, trained on computer (topology optimization), natural (leaf microstructures), and human-generated parametric (constructal) designs, used a variational autoencoder (VAE) \cite{kingma2014autoencoding} to map area-to-point designs into a compact latent space.
New dendritic configurations were created by sampling and decoding, with the latent space size controlling the design freedom-to-morph, where larger spaces enabled more detailed structures.
Separately, an unsupervised diffusion-based model \cite{ddpm2020} trained on the same dataset was used, enabling greater freedom-to-morph by producing diverse, asymmetric dendritic structures. 
An evolutionary algorithm then navigated the generative design space, systematically enhancing diversity by combining advantageous traits and balancing exploration with exploitation. 
This process led to the evolution of efficient, hierarchical tree networks, exhibiting a higher degree of morphological diversity than those designed through traditional parametric approaches, as presented in Fig. \ref{fig:atp_samples}.

\begin{figure}[h!]
    \centering
    \includegraphics[width=0.8\textwidth]{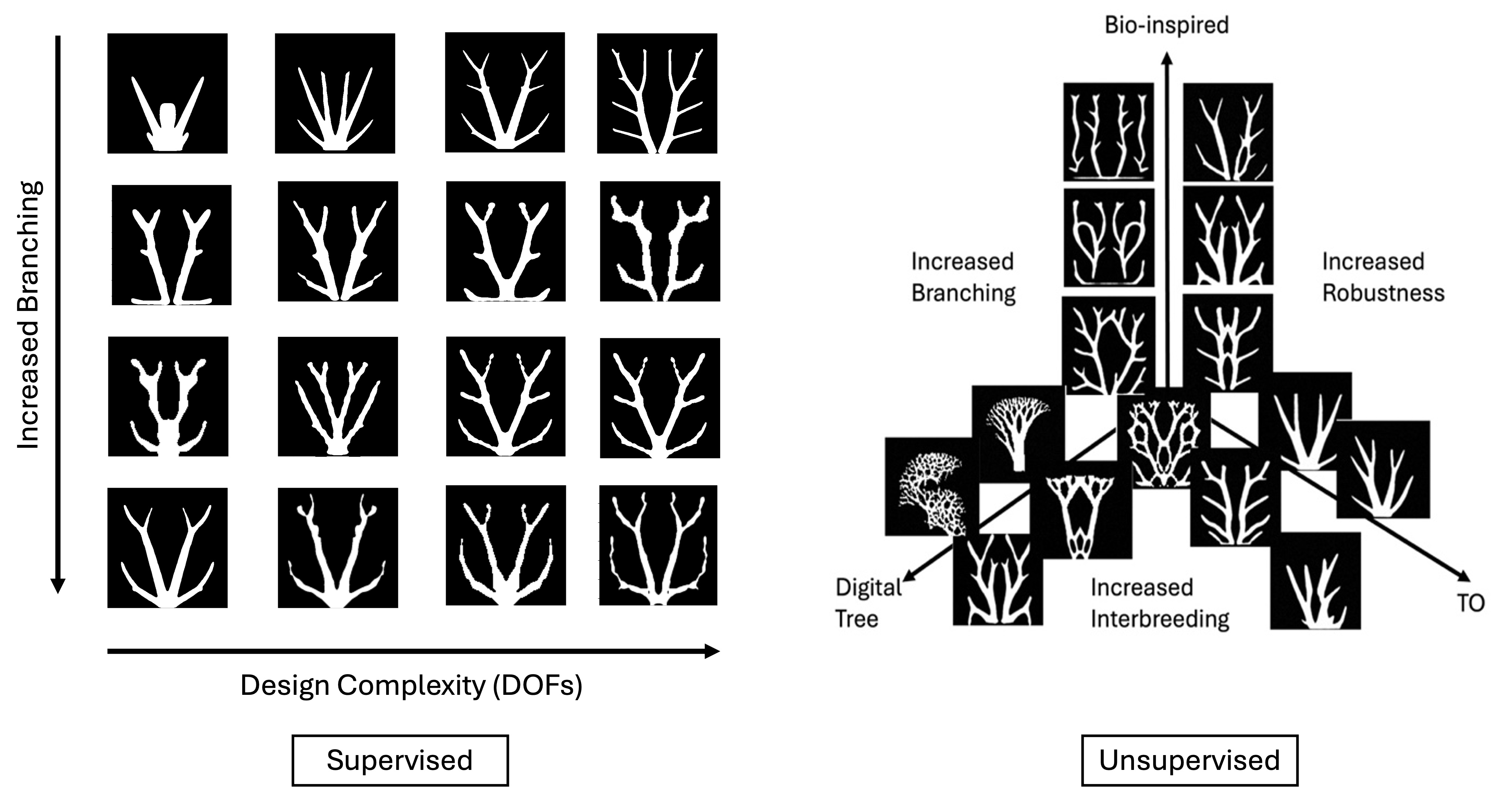}
    \caption{\textbf{Generative Design Solutions for Area-to-Point Systems.} Example configurations of networks high-conductivity material optimized for the area-to-point conduction problem using generative models. The designs, created through both supervised and unsupervised approaches, demonstrate diverse, hierarchical structures that improve flow efficiency compared to traditional parametric designs.}
    \label{fig:atp_samples}
\end{figure}

To create a complete vascular flow network for a cold-plate, two area-to-point (ATP) unit cells, generated as described, were combined in a canopy-to-canopy configuration. 
One unit cell acts as the supply tree, channeling coolant from the inlet to the central battery area. 
The second, acting as the collection tree, then gathers coolant from the area and directs it to the outlet. 
This canopy-to-canopy approach, previously used for parametric constructal design \cite{}, is now leveraged to generate hierarchical dendritic structures with complex features that reduce combined thermal and fluid losses. 
The result is a blueprint of a vascular flow path, with channels of variable size, that maximizes coolant-battery thermal contact and minimizes pumping losses between the fluid source and sink. 

To further optimize this design, we employ a multi-step evolutionary algorithm. 
Initially, we use the Non-dominated Sorting Genetic Algorithm II (NSGA-II) to quickly improve the early stages of the flow network shape, assuming perfect thermal contact and uniform heat generation. 
This step allows for broad design space exploration with reduced computational costs. 
The result is a non-dominated Pareto front, which shows the trade-off between the competing objectives of minimizing pumping losses and thermal resistance.
Subsequently, each high-performing design is evolved in parallel along paths of low resistance, as determined by the gradient of the adjoint sensitivity field with respect to the design variable, similar to how conventional gradient-based topology optimization iteratively adjusts material density based on sensitivity analysis.
In this second phase of optimization, the full physics, including conjugate heat transfer and bi-directional thermal-electrochemical coupling, is considered.

Traditional methods such as material density topology optimization (TO) start from uniform or parametric material distributions and subsequently rely on gradient-based sensitivity analysis for flow path optimization.
The inherent reliance on gradient information not only hinders adaptability across diverse physical systems but also restricts design diversity, limiting the range of achievable trade-offs within a given design space. 
Our method overcomes this by employing a transfer learning approach. 
By initially sampling from the previously learned latent space of the area-to-point (ATP) designs, we leverage the a priori knowledge of efficient flow configurations identified by natural and computational systems. 
This enables the navigation complex design landscapes with significantly fewer iterations. 
Starting the gradient-based evolution stage from a well-formed canopy-to-canopy network is particularly relevant in this case, considering the computational requirements of the high-fidelity coupled model used.
Moreover, this secondary process can be parallelized across each candidate in the Pareto front, enabling the exploration of the thermal and fluid trade-offs within a single optimization loop.

\begin{figure}[h!]
    \centering
    \includegraphics[width=0.75\textwidth]{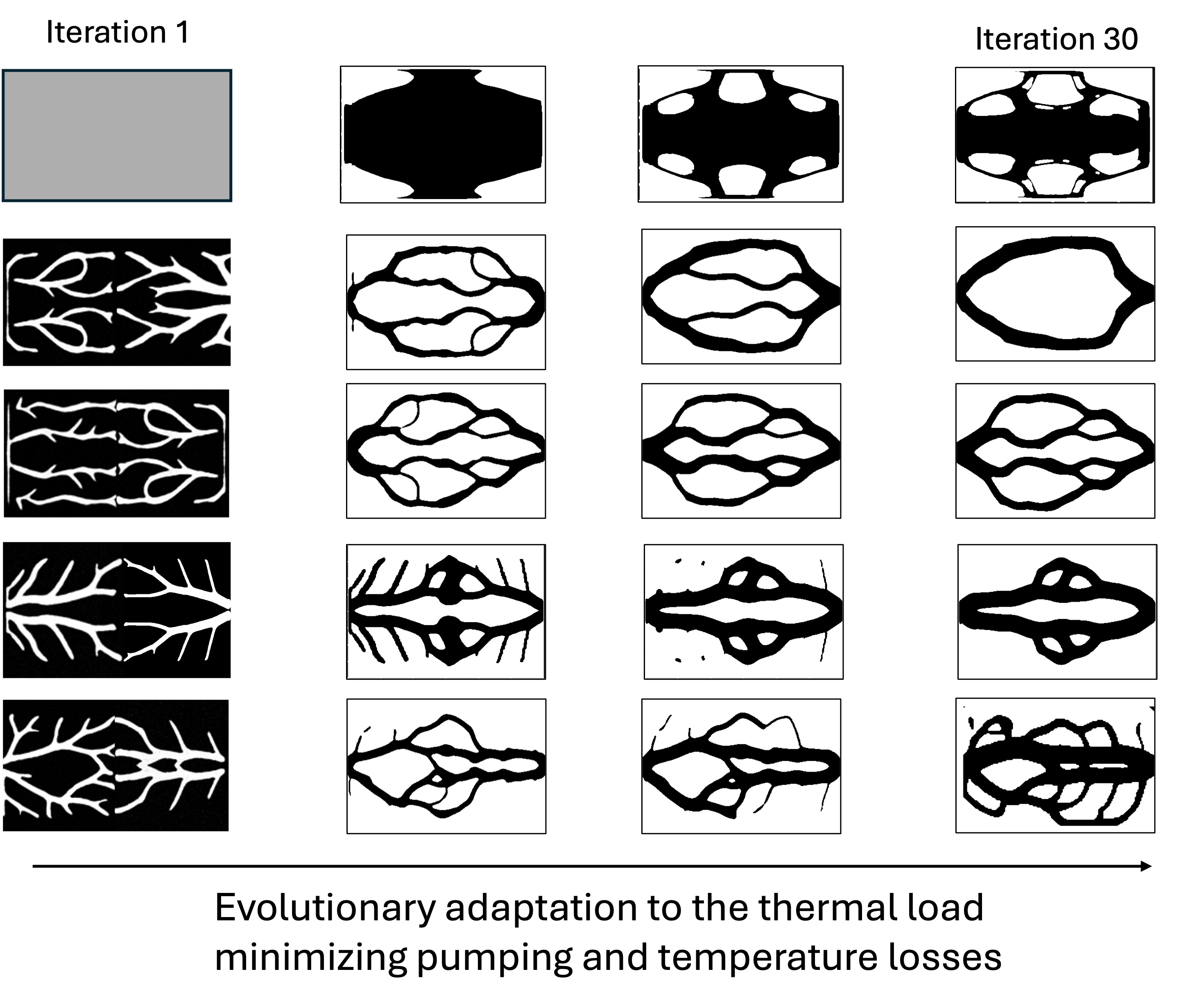}
    \caption{\textbf{Generative Design of Vascular Flow Networks.} ATP substructures (Fig. \ref{fig:atp_samples}) are combined into a single inlet -- single outlet network and subjected to evolutionary optimization. Loading-in the genetic content from the tree-like conduction case dramatically speeds-up the optimization process.}
    \label{fig:vascular_design}
\end{figure}

\begin{figure}[h!]
    \centering
    \includegraphics[width=0.85\textwidth]{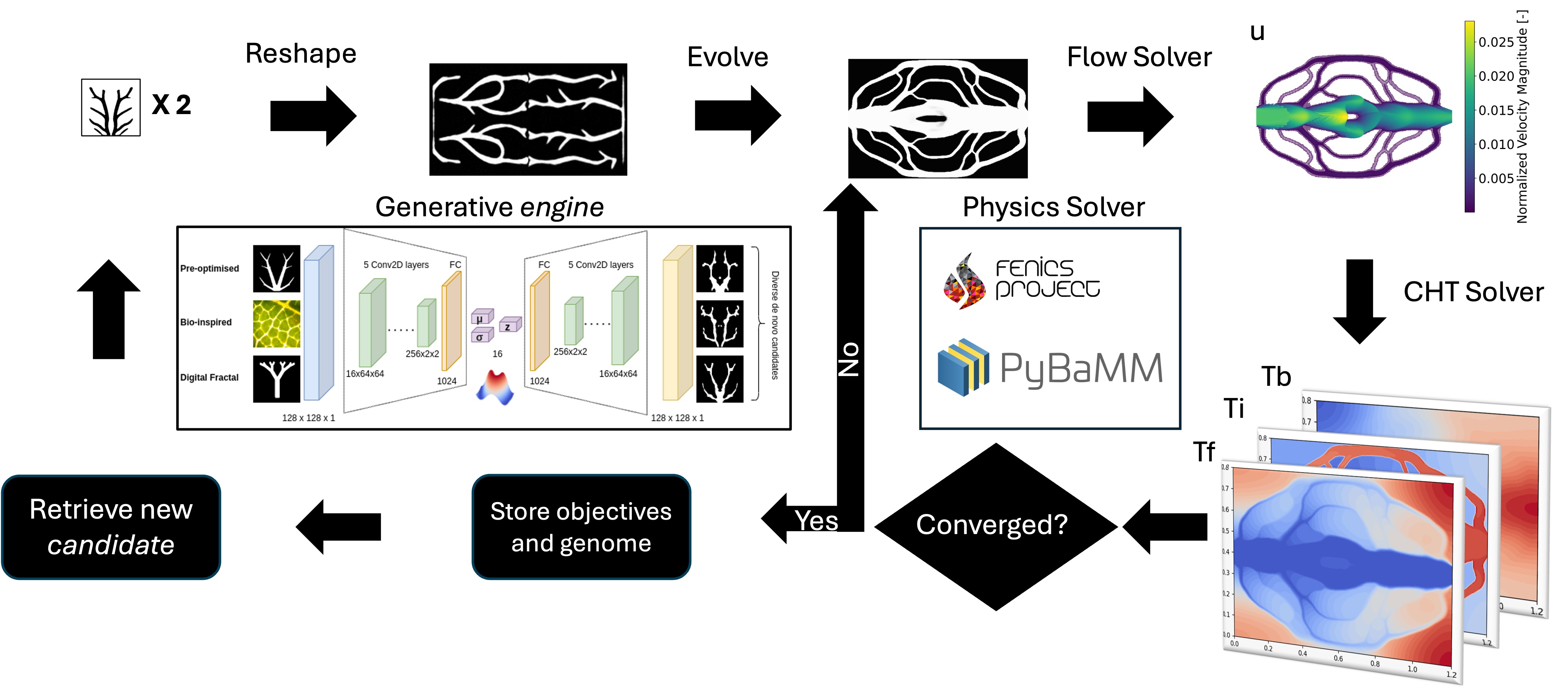}
    \caption{\textbf{Evolutionary Optimization Flowchart.} Vascular flow networks, generated by the generative engine from two canopy-to-canopy ATP structures, undergo evolution via the globally convergent method of moving asymptotes (GCMMA). Each design is evaluated using a physics solver to compute thermal and fluidic losses. NSGA-II is then employed to identify the Pareto front of the multi-objective trade-off between thermofluid performance metrics.}
    \label{fig:evolutionary_opt}
\end{figure}

\subsection{Coupled Thermal-Electrochemical Model}
The thermal-electrochemical model of the pouch cell is constructed using PyBaMM’s \cite{Pybamm} higher-dimensional pouch cell framework derived using asymptotic analysis \cite{Timms2021}.
This "2+1D" x-lumped model assumes negligible temperature variation through the cell thickness while solving for the the spatial temperature distribution on the battery surface $T_b(y,z,t)$. 
The governing partial differential equation accounts for heat conduction, internal heat generation, and convective cooling at the surfaces, parameterized by effective thermal conductivity $\lambda_{eff}$ and heat transfer coefficients at the current collectors, tabs, and edges. 
This approach enables the accurate dynamic coupling of thermal and electrochemical effects while maintaining computational efficiency \cite{Marquis2019AnAD}.

\begin{equation}
\begin{cases}
\rho_{eff} \frac{\partial T(y,z,t)}{\partial t} = \lambda_{eff} \nabla^2 T(y,z,t) + \overline{Q}(y,z,t) - \dfrac{(h_{cn}+h_{cp})A}{V} \big( T(y,z,t)-T_{\infty}(y,z,t) \big), \\[10pt]

-\lambda_{eff} \nabla T \cdot \boldsymbol{n} = \dfrac{L_{cn}h_{cn}}{L_{cn}+L_n+L_s+L_p+L_{cp}} \big( T - T_\infty \big), \quad &\text{(Negative tab)} \\[10pt]

-\lambda_{eff} \nabla T \cdot \boldsymbol{n} = \dfrac{L_{cp}h_{cp}}{L_{cn}+L_n+L_s+L_p+L_{cp}} \big( T - T_\infty \big), \quad &\text{(Positive tab)} \\[10pt]

-\lambda_{eff} \nabla T \cdot \boldsymbol{n} = 0, \quad &\text{(Elsewhere)}
\end{cases}
\end{equation}

The variables in the equations are defined as follows:

\begin{tabbing}
    \hspace{2cm} \= \hspace{12cm} \= \\ % Adjust spacing as needed
    \textbf{Symbol} \> \textbf{Description} \> \textbf{Units} \\[5pt]
    $T(y,z,t)$ \> Temperature distribution within the pouch cell at position $(y,z)$ and time $t$ \> K \\
    $\rho_{eff}$ \> Effective volumetric heat capacity of the cell material \> $\text{J} \cdot \text{m}^{-3} \cdot \text{K}^{-1}$ \\
    $\lambda_{eff}$ \> Effective thermal conductivity of the cell material \> $\text{W} \cdot \text{m}^{-1} \cdot \text{K}^{-1}$ \\
    $\overline{Q}(y,z,t)$ \> Volumetric heat source term due to electrochemical and resistive heating \> $\text{W} \cdot \text{m}^{-3}$ \\
    $A$ \> Surface area of the pouch cell \> $\text{m}^2$ \\
    $V$ \> Volume of the pouch cell \> $\text{m}^3$ \\
    $h_{cn}$ \> Heat transfer coefficient at the negative current collector surface \> $\text{W} \cdot \text{m}^{-2} \cdot \text{K}^{-1}$ \\
    $h_{cp}$ \> Heat transfer coefficient at the positive current collector surface \> $\text{W} \cdot \text{m}^{-2} \cdot \text{K}^{-1}$ \\
    $T_{\infty}(y,z,t)$ \> Ambient (cold plate) temperature at position $(y,z)$ and time $t$ \> K \\
    $L_{cn}$ \> Length of the negative current collector tab \> m \\
    $L_{cp}$ \> Length of the positive current collector tab \> m \\
    $L_n, L_s, L_p$ \> Thicknesses of the negative electrode, separator, and positive electrode \> m \\
    $\boldsymbol{n}$ \> Outward unit normal vector at the boundary of the domain \> -- \\
\end{tabbing}

In this setup, the edge of the pouch cell and the [positive/negative] electrode side of the pouch cell are assumed to be insulated to represent a realistic integration of the cell in a battery pack.
The heat source term \( \overline{Q} (y,z,t) \) accounts for multiple heating mechanisms and is expressed as  \[
\overline{Q} = Q_{Ohm,k} + Q_{rxn,k} + Q_{rev,k} + Q_{cr}
\]
where \( Q_{Ohm,k} \) represents Ohmic heating due to resistive losses in both the solid and electrolyte phases, while \( Q_{rxn,k} \) corresponds to irreversible heating generated by electrochemical reactions. Additionally, \( Q_{rev,k} \) captures reversible heating effects arising from entropy changes during charge and discharge cycles, and \( Q_{cr} \) accounts for heating due to contact resistance at interfaces within the cell. This model provides a detailed representation of the thermal behaviour of the pouch cell, enabling accurate prediction of temperature evolution during operation \cite{Pybamm, Timms2021, Marquis2019AnAD}.

The electrochemical model of the lithium-ion battery can be selected from several available options in PyBaMM, each offering different levels of complexity and accuracy. These options include:

\begin{itemize}
    \item \textbf{SPM (Single Particle Model)} and \textbf{SPMe (Single Particle Model with Electrolyte)} \cite{SPM}: These models assume a uniform concentration profile across the electrodes and provide a simplified representation of electrochemical processes, with SPMe additionally accounting for the electrolyte phase.
    \item \textbf{MPM (Multi-Particle Model)} \cite{MPM}: This model improves upon the SPM by dividing the electrodes into multiple particles, allowing for the capture of concentration gradients and more detailed internal processes.
    \item \textbf{DFN (Doyle-Fuller-Newman Model)} \cite{DFN}: This model incorporates spatial concentration gradients in both the electrolyte and electrodes, offering a detailed representation of electrochemical behaviour and transport processes.
    \item \textbf{NT (Newman-Tobias Model)} \cite{Newman_1962, Chu2020}: Based on the assumption of a uniform concentration profile in the electrolyte, this model includes nonlinear Butler-Volmer kinetics and tracks the average concentration in the solid-phase electrodes, with an optional inclusion of mass transport in individual particles.
\end{itemize}

To investigate the influence of mesh resolution on the solution, we discretize the electrochemical and thermal domains, denoted as \textit{EC} and \textit{HT}, respectively. 
For the electrochemical model, the discretization involves parameters such as $x_n$, $x_s$, $x_p$, which represent the spatial discretization of the negative electrode, separator, and positive electrode. 
Similarly, the thermal discretization involves the dimensions $y$ and $z$, which are associated with the spatial resolution of the heat generation and the temperature distribution across the surface of the battery.

By solving the model with varying mesh resolutions and comparing the results, we can determine the optimal model and mesh configuration for accurate simulations that balance computational cost with accuracy.
The findings of this convergence study are summarized in Fig. \ref{fig:benchmark}.
Among the configurations tested, $\text{SPMe\_EC4\_HT16}$ is found to be the optimal choice. 
This configuration produces a solution that is within 2\% of the maximum heat generation obtained from the more computationally expensive DFN model, while overpredicting the mean heat generation by up to 10\%. 
Despite the slight overprediction, which would also result in conservative estimates, the $\text{SPMe\_EC4\_HT16}$ configuration provides a substantial computational benefit, solving the problem up to 300 times faster than the DFN model.
This balance of accuracy and speed makes $\text{SPMe\_EC4\_HT16}$ a suitable choice for this study, given the large number of iterations required for the evolutionary design algorithm.

\begin{figure}[h!]
    \centering
    \includegraphics[width=0.75\textwidth]{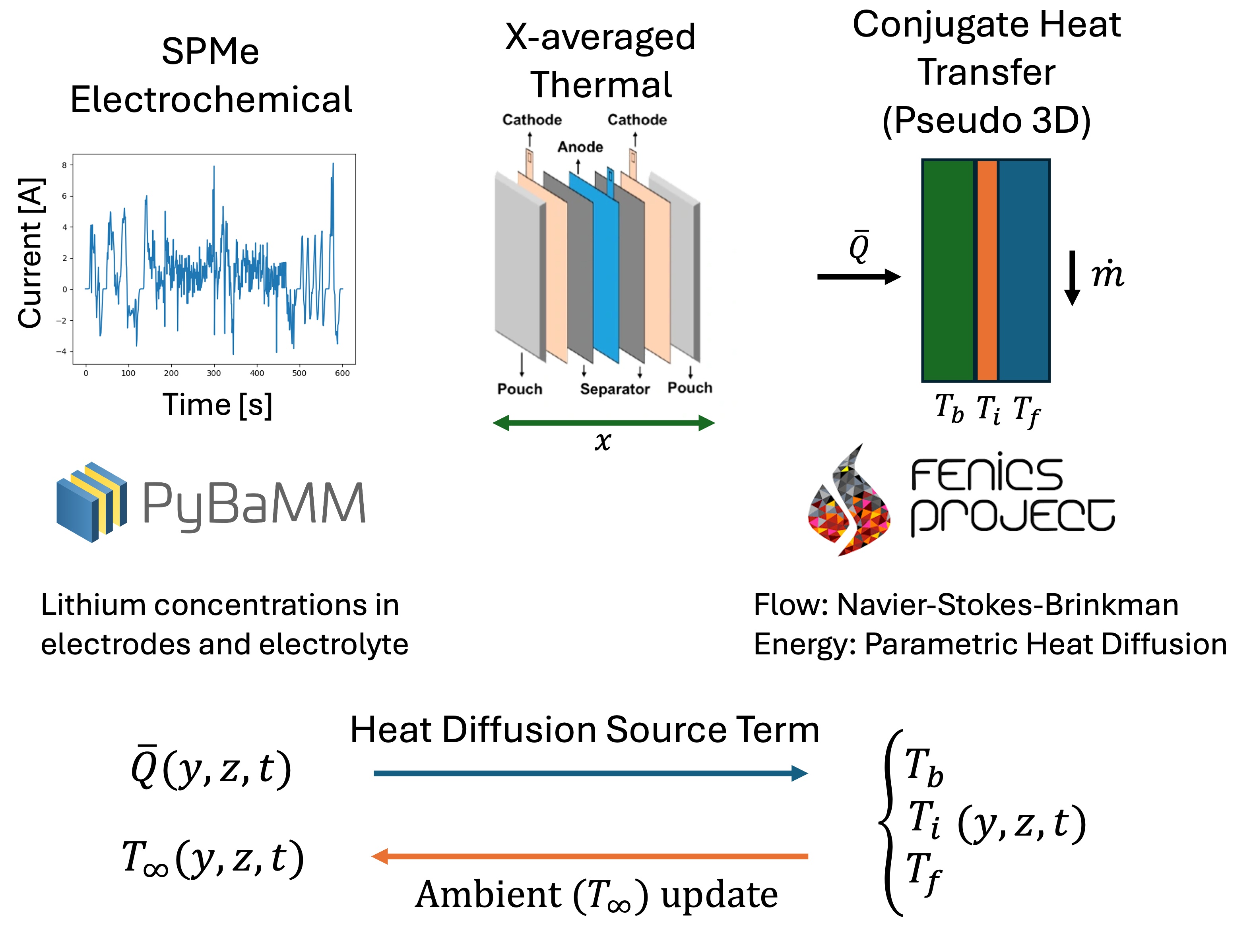}
    \caption{\textbf{Coupled Thermal-Electrochemical Approach using PyBaMM and FEniCS.} 
    The electrochemical behaviour of a battery cell is simulated using PyBaMM, which takes the current profile as an input and provides the volumetric heat generation. This information is then coupled with a conjugate heat transfer model implemented in FEniCS, where the temperature distribution and coolant flow are solved. The ambient temperature is updated in PyBaMM every 8th time step, representing the thermal back coupling.}
    \label{fig:atp_samples}
\end{figure}

\begin{figure}[H]
    \centering
    \includegraphics[width=0.9\textwidth]{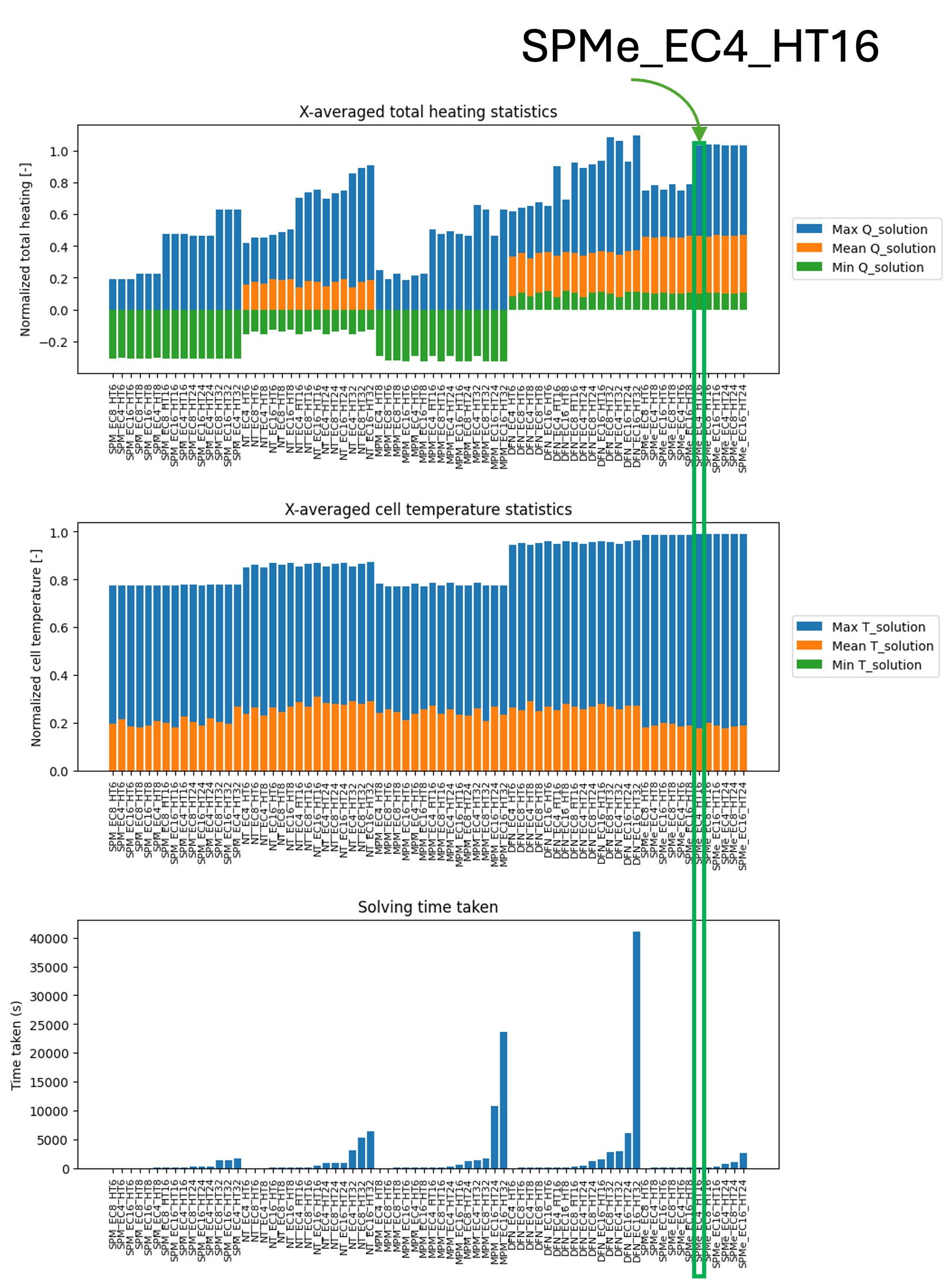}
    \caption{\textbf{Thermal-Electrochemical Model Convergence Study.} The effect of mesh resolution on the thermal-electrochemical model solution is investigated by comparing various Li-ion models (SPM \cite{SPM}, SPMe \cite{SPM}, MPM \cite{MPM}, DFN \cite{DFN}, NT \cite{Newman_1962, Chu2020}) and thermal discretization configurations. Among the tested configurations, $\text{SPMe\_EC4\_HT16}$ is found to produce accurate results at a fraction of computational cost of the DFN model.}
    \label{fig:benchmark}
\end{figure}

\subsection{Pseudo-3D Conjugate Heat Transfer Model}
\label{sec:pseudo3d_cht}

Heat generated in the pouch cell passes through the battery--cold-plate interface and the cold-plate structure before it is removed by the coolant. The battery model described in Section~2.2 supplies a spatially and temporally varying electrochemical heat-generation field and resolves the in-plane cell temperature. The cold-plate model resolves the in-plane flow and thermal response of the channel network. Coupling these descriptions allows the design objectives to be evaluated at the battery surface, rather than inferred from coolant temperature alone.

The model is termed pseudo-3D because temperature and flow variations are resolved in the cell plane, whereas transport through the thickness of the battery, thermal interface material (TIM), and cold plate is represented by effective thermal conductances. Let $(y,z)$ denote coordinates in the cell plane, $T_b$ the battery temperature, $T_i$ the TIM-layer temperature, and $T_f$ the temperature of the effective cold-plate domain, which represents the coolant and solid regions through the design variable $\gamma$. The layer-averaged thermal balances take the form
\begin{equation}
\begin{aligned}
C_b\frac{\partial T_b}{\partial t}
-\nabla_{\parallel}\cdot\bigl(K_b\nabla_{\parallel}T_b\bigr)
&=q_b''(y,z,t)-G_{bi}(T_b-T_i),\\
-\nabla_{\parallel}\cdot\bigl(K_i\nabla_{\parallel}T_i\bigr)
&=G_{bi}(T_b-T_i)-G_{if}(\gamma)(T_i-T_f),\\
-\nabla_{\parallel}\cdot\bigl(K_{\mathrm{eff}}(\gamma)\nabla_{\parallel}T_f\bigr)
+C_{f,\mathrm{eff}}(\gamma)\,\boldsymbol{u}\cdot\nabla_{\parallel}T_f
&=G_{if}(\gamma)(T_i-T_f).
\end{aligned}
\label{eq:pseudo3d_thermal}
\end{equation}
Here $\nabla_{\parallel}$ acts in the cell plane; $q_b''=d_b Q$ is the battery heat-generation rate per unit cell area corresponding to a thickness-averaged volumetric source $Q$ and represented thickness $d_b$. The areal battery heat capacity is $C_b=(\rho c_p)_b d_b$. The in-plane conductances are $K_b=\lambda_b d_b$, $K_i=\lambda_i d_i$, and $K_{\mathrm{eff}}=\lambda_{\mathrm{eff}}d_{CP}$, where $d_i$ and $d_{CP}$ are the represented TIM and cold-plate thicknesses. Similarly, $C_{f,\mathrm{eff}}=(\rho c_p)_{\mathrm{eff}}d_{CP}$ is the effective areal heat capacity in the cold-plate domain. The battery retains its transient response; the TIM and cold-plate balances describe their thermal states for the applied load at each coupling exchange. The battery's tab and external-boundary conditions follow the pouch-cell formulation in Section~2.2.

The interlayer conductances $G_{bi}$ and $G_{if}$ (in $\mathrm{W\,m^{-2}\,K^{-1}}$) describe transport across the battery--TIM and TIM--cold-plate links. The TIM's nominal resistance per unit area is $R_{\mathrm{TIM}}''=d_i/\lambda_i$; battery-side and plate-side contact resistances can be included in the corresponding links. The individual resistances must be assigned consistently so that TIM conduction is not counted twice. The paired interlayer terms in Eq.~\eqref{eq:pseudo3d_thermal} have equal magnitude and opposite sign in adjacent layers, enforcing heat conservation across each represented interface. The fluid-side contribution to $G_{if}$ depends on the local channel state and heat-transfer coefficient; consequently, the battery--coolant temperature difference is not represented by a single prescribed coolant temperature.

The design variable $\gamma\in[0,1]$ denotes the local material state, with $\gamma=0$ corresponding to solid and $\gamma=1$ to fluid. The effective in-plane conductivity is interpolated between its solid and fluid limits as
\begin{equation}
\lambda_{\mathrm{eff}}(\gamma)
=\lambda_s+(\lambda_f-\lambda_s)
\frac{\gamma(1+p_a)}{\gamma+p_a},
\label{eq:lambda_interpolation}
\end{equation}
where $p_a>0$ controls the interpolation. The effective heat capacity and interlayer conductance are likewise evaluated according to the local material state. The interpolated coolant-side heat-transfer coefficient is
\begin{equation}
h(\gamma)=h_f+(h_s-h_f)\frac{1-\gamma}{1+q_a\gamma},
\label{eq:h_interpolation}
\end{equation}
where $h_s$ and $h_f$ are the coefficients at the solid and fluid endpoints and $q_a$ is the interpolation parameter. This coefficient contributes to the effective interlayer conductance; it does not replace the battery, TIM, and contact resistances.

\begin{figure}[!htbp]
    \centering
    \includegraphics[width=0.9\textwidth]{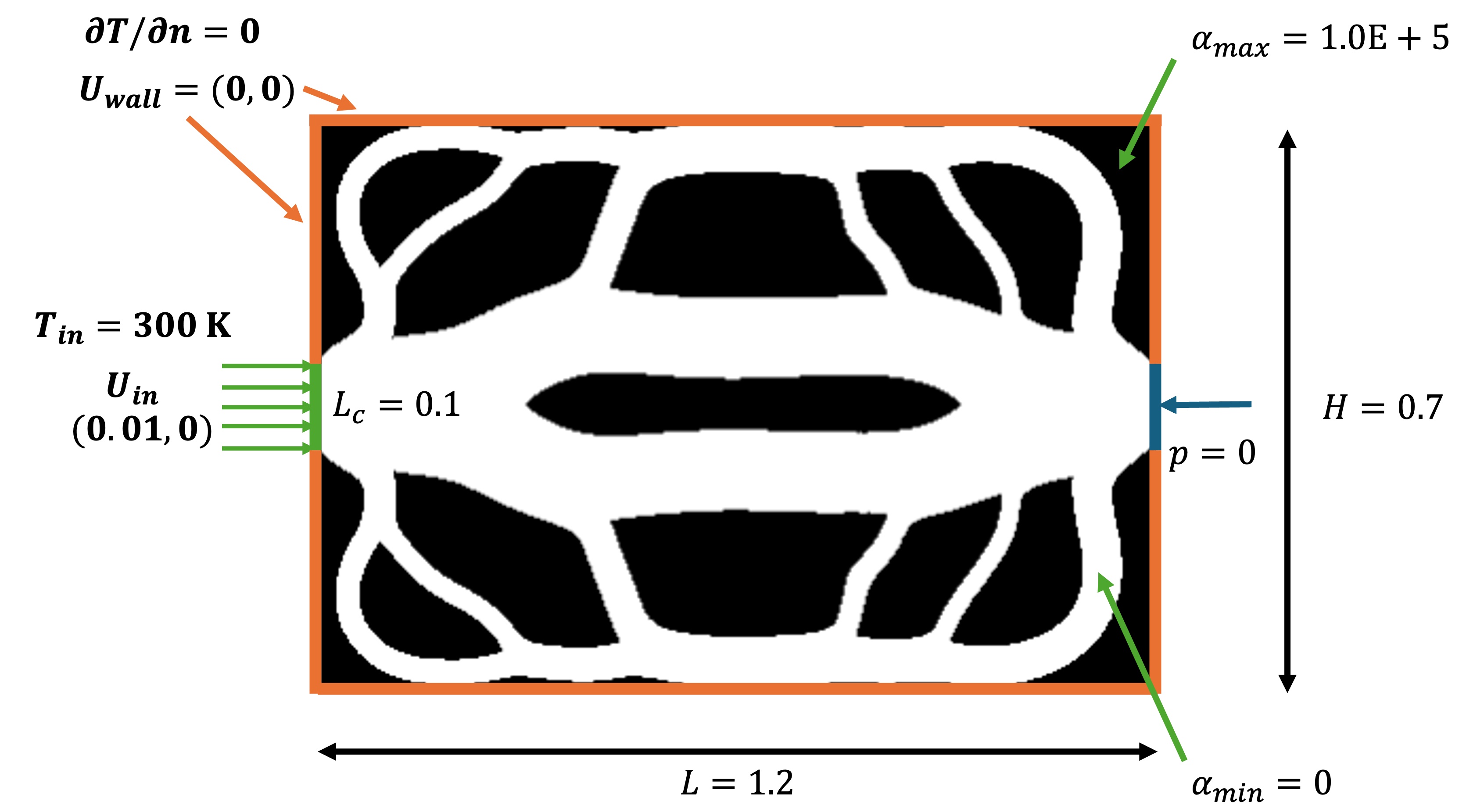}
    \caption{In-plane cold-plate design domain for the Navier--Stokes--Brinkman calculation, showing inlet, outlet, wall, and thermal boundary conditions. The design variable $\gamma$ interpolates between solid- and fluid-dominated regions; dimensions and inlet values shown correspond to the illustrated computational case.}
    \label{fig:domain}
\end{figure}

Coolant motion in the design domain is described by the incompressible Navier--Stokes--Brinkman equations,
\begin{equation}
\nabla_{\parallel}\cdot\boldsymbol{u}=0,\qquad
\rho_f(\boldsymbol{u}\cdot\nabla_{\parallel})\boldsymbol{u}
=-\nabla_{\parallel}p+\mu_f\nabla_{\parallel}^{2}\boldsymbol{u}
-\alpha(\gamma)\boldsymbol{u},
\label{eq:ns_brinkman}
\end{equation}
where $\boldsymbol{u}$ is the in-plane velocity, $p$ is pressure, and $\rho_f$ and $\mu_f$ are the coolant density and dynamic viscosity. The Brinkman resistance is interpolated as
\begin{equation}
\alpha(\gamma)=\alpha_f+(\alpha_s-\alpha_f)
\frac{1-\gamma}{1+q_a\gamma},\qquad \alpha_s>\alpha_f\geq 0.
\label{eq:alpha_interpolation}
\end{equation}
The resistance is therefore largest in solid-dominated regions, and its force opposes the local velocity. Inlet velocity and temperature are prescribed, outlet pressure is set to zero gauge, impermeable boundaries have no slip, and the external lateral thermal boundaries are taken as adiabatic. Figure~\ref{fig:domain} illustrates the computational domain and the boundary conditions for the depicted design case.

Coupling proceeds by transferring the heat-generation field calculated in PyBaMM to the battery thermal balance and returning the interface temperature calculated by the thermal solver as a spatially resolved boundary condition for PyBaMM. The cold-plate velocity and temperature fields are evaluated for the resulting load. In the present implementation, the temperature boundary field provided to PyBaMM is updated every eighth electrochemical time step. The exchange links local cooling to the subsequent temperature-dependent electrochemical response without explicitly meshing the full three-dimensional assembly. Because the electrochemical and thermal properties of the commercial test cell were not fully identified, the model is used to compare design responses and spatial thermal trends, not to claim cell-specific prediction of its absolute temperature history.

\subsection{Thermal Design Objectives}
Effective thermal management in battery-cold plate systems requires optimizing both cooling efficiency and energy consumption. 
We define three primary objectives, one concerning the pumping power required to drive the system and two concerning the temperature of the battery surface.

Pumping Losses represent the energy dissipated due to the coolant flow through the cold plate. 
These losses are characterized by an integral based on the velocity field and the Brinkman term, which accounts for both the viscous and inertial contributions to the flow resistance. 
The pumping losses are quantified as:
\begin{equation}
    \Psi_{f} = \int_{\Omega} \left( \frac{1}{2}\mu \sum_{i,j} \left( \frac{\partial u_i}{\partial x_j} + \frac{\partial u_j}{\partial x_i} \right) + \sum_i \alpha(x) \, u_i \cdot u_i \right) d\Omega,
\end{equation}
where $\mu$ is the dynamic viscosity, $\alpha(x)$ is a term related to the flow resistance through the porous medium, and $\mathbf{u}$ is the velocity vector. 
Minimizing $\Psi_{f}$ reduces auxiliary power consumption, at a penalty in overall cooling performance.

Mean Battery Surface Temperature represents the overall thermal efficiency of the system, characterized by the average temperature across the battery surface. 
It is expressed mathematically as:
\begin{equation}
    T_{\text{mean}} = \frac{1}{\Omega} \int_{\Omega} T_b \, d\Omega,
\end{equation}
where $\Omega$ is the battery surface area and $T_b$ is the local surface temperature. 
Lowering $T_{\text{mean}}$ improves the thermal management of the system, leading to better performance and a longer battery lifespan.

Hotspot Mitigation focuses on reducing localized high temperatures on the battery surface, which are critical for ensuring the longevity and efficiency of the battery. 
The severity of hotspot formation is quantified using an Lp-norm of coefficient $p=10$, which is defined as:
\begin{equation}
    T_{\text{hot}} =  \frac{1}{\Omega} \left( \int_{\Omega} |T - T_{\text{mean}}|^p \right)^{1/p} \, d\Omega,
\end{equation}
where $\Omega$ is the battery surface area, $T$ is the temperature at a given point, and $T_{\text{mean}}$ represents the mean temperature observed of the battery surface. 
Minimizing $T_{\text{hot}}$ reduces hotspots, preventing thermal degradation and ensuring better performance over time.

The evaluation of these objectives is performed under both steady-state and time-dependent conditions. 
While steady-state analysis provides a snapshot of the system performance at equilibrium, time-dependent analysis, especially in the context of a given drive cycle, introduces additional complexity by considering dynamic temperature variations. 
To address this, we incorporate a time-variance objective that evaluates the system performance over time. 
The time-variance is quantified by the following equation, which calculates the variance ($\sigma^2$) of the maximum temperature over a given time period:
\begin{equation}
    \sigma^2 = \frac{1}{\tau} \int_0^\tau \left( T_{\text{max}}(t) - \overline{T_{\text{max}}} \right)^2 \, \mathrm{d}t,
\end{equation}
where $T_{\text{max}}(t)$ is the maximum temperature at time $t$, $\overline{T}_{\text{max}}$ is the time-averaged maximum temperature over the drive cycle period $\tau$. 
This analysis allows for a more granular understanding of thermal management throughout varying operational conditions, enabling optimization of both cooling and energy consumption strategies in response to fluctuating thermal loads.

To make the trade-off between competing objectives, we utilize entropy generation minimization. 
This approach identifies the design that minimizes entropy production, balancing the conflicting demands of cooling efficiency and energy consumption. 
The number of entropy production units, \( N_s \), can be expressed as the sum of two terms: the first term represents the entropy generation due to frictional losses, involving mass flow rate \( \dot{m} \), density \( \rho \), heat flux \( q' \), and the pressure gradient \( dP / {dx} \), while the second term accounts for thermal irreversibility, involving the temperature difference between the source and the fluid \( \Delta T \) and the fluid temperature \( T \) \cite{Bejan1978}. 
The formula for the total entropy generation is given as:
\begin{equation}
    N_s = \frac{\dot{m}}{\rho q'} \left(- \nabla P \right) + \frac{\Delta T}{T} \left( 1+\frac{\Delta T}{T} \right) ^ {-1}
\end{equation}

\subsection{Flow Network Reconstruction with Straight Channels}
An computer vision algorithm was developed to convert the \textit{organic} flow networks developed into a structured arrangements of straight channels, enhancing manufacturing efficiency.
The process begins with image preprocessing, which includes noise reduction through median filtering and morphological operations. 
Median filtering removes small-scale noise while preserving edges, ensuring that fine details of the network are retained. 
Morphological operations, such as opening and closing, smooth out irregularities by removing small protrusions and filling gaps \cite{scikit-image}.

Skeletonization reduces the network to a single-pixel-wide structure, preserving its topological properties while eliminating excess width. 
This is achieved using iterative thinning algorithms and distance transforms, ensuring that the network maintains connectivity while simplifying its representation.
To this end, we use the \texttt{skeletonize} function from the \texttt{skimage} library in Python.

\begin{figure}
    \centering
    \includegraphics[width=\textwidth]{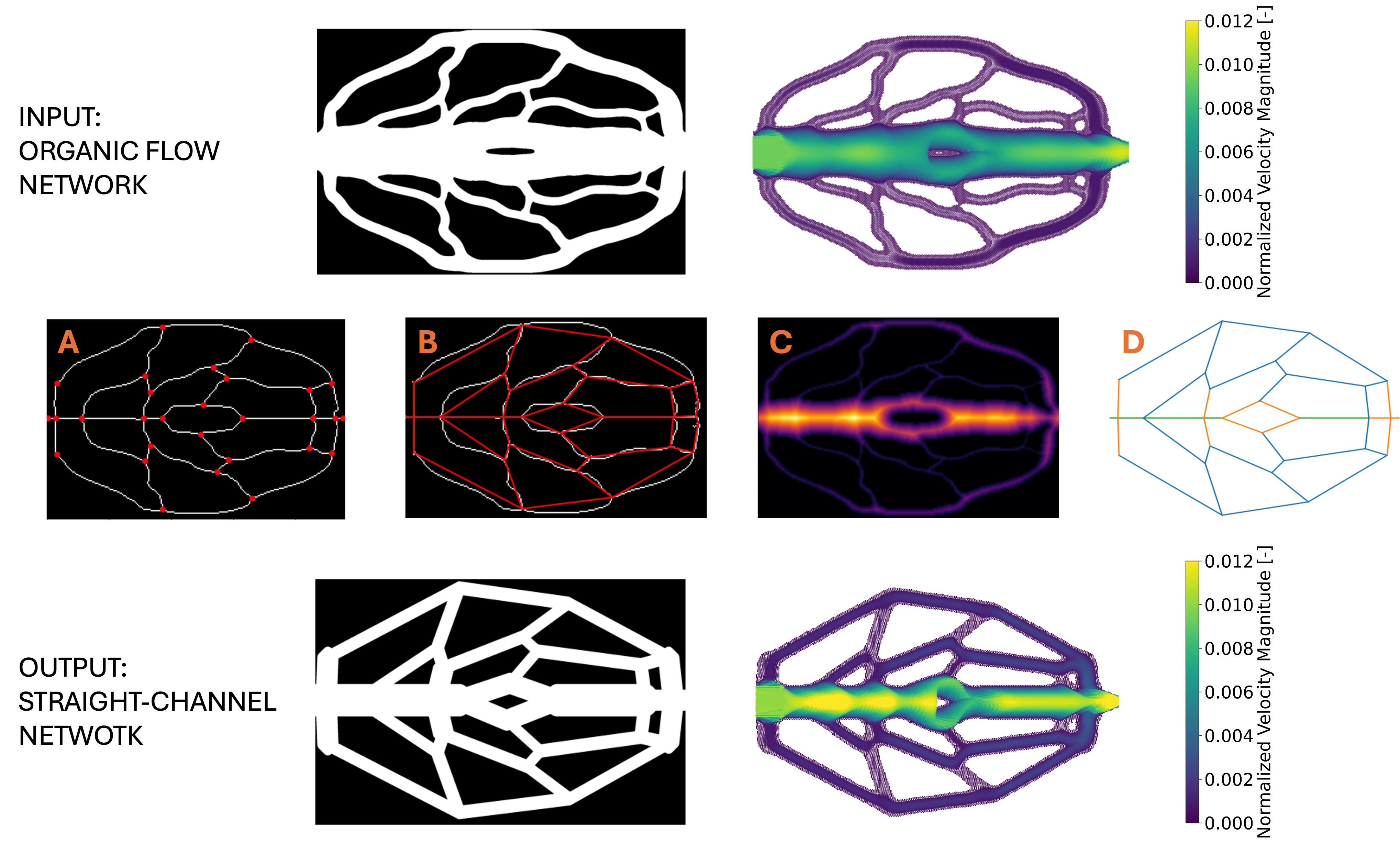}
    \caption{Comparison of the organic flow network (top) and its structured reconstruction (bottom). (A) Skeletonized representation of the original network, highlighting flow pathways and junctions. (B) Graph representation with identified branch points and channel segments. (C) Euclidean Distance Transform (EDT) map illustrating channel thickness distribution. (D) CNC tool path colour-coded by thickness, showing the mapped manufacturing strategy for varying channel widths.}
    \label{fig:reconstruction}
\end{figure}

Further, the branch points are identified to determine intersections within the flow network, and the skeletonized structure is converted into a graph representation. 
Nodes correspond to branch points and endpoints, while edges represent channel segments, as seen in Fig. \ref{fig:reconstruction}A. 
This graph structure encodes connectivity and flow paths, allowing for precise geometric modifications.

Channel thickness is computed using the Euclidean Distance Transform (EDT), which assigns each pixel a value corresponding to its distance from the nearest background pixel. 
This provides a local measure of channel width, ensuring that the reconstructed straight channels maintain the average thickness of the original organic structure, preserving flow characteristics.
Inset C in Fig. \ref{fig:reconstruction}C shows a distance map of the \textit{organic} flow network received as an input.

Breadth-first search (BFS, \cite{Kozen1992}) is employed to analyze direct connections between branch points, simulating the trajectory of a fluid particle from the inlet node to the outlet node in the graph. 
By traversing the network layer by layer, BFS ensures that the shortest and most direct paths are identified, facilitating the accurate reconstruction of structured channels.

To optimize manufacturability, k-means clustering is applied to the thickness values extracted from the EDT matrix \cite{scikit-image}. 
Three representative thickness values are selected based on the histogram distribution, reducing the complexity of machining while preserving essential morphological characteristics. 
This step ensures that the structured network remains faithful to the original flow geometry while significantly simplifying fabrication.

Volume fraction control is implemented to fine-tune the channel thicknesses, ensuring that the final structure achieves the desired porosity, defined as the ratio of fluid to solid phases in the design domain.
The resulting network consists of straight channels with three standardized thickness values, each corresponding to available end-effectors in the CNC toolbox. 
This design choice minimizes machining time while maintaining morphological similarity and flow performance.

Crucially, the main features of the original flow network—such as primary pathways and junction connectivity -- are preserved in the reconstructed structure, ensuring that the overall flow distribution remains consistent. 
As shown in Fig. \ref{fig:reconstruction}, the reconstructed network exhibits a similar flow profile to the organic design, demonstrating that the essential transport characteristics are maintained. 
However, due to the geometric constraints imposed by straight channels, more abrupt changes in flow velocity are observed, particularly at branch points where sudden expansions or contractions occur. 
Despite these localized variations, the structured network effectively balances manufacturability with fluidic losses, providing a viable alternative for streamlining fabrication while retaining key functional attributes.

The plate tested in this study was machined from the selected organic flow-network design. Figure~\ref{fig:experimental_assembly} shows the design drawing, the as-built machined organic network, and its integration with the pouch cell and thermal interface. Accordingly, the experimental results characterize the machined organic plate, whereas the reconstructed and parametric layouts serve as numerical comparators.

\subsection{Experimental Setup and Measurement Strategy}
\label{sec:experimental_methods}

The experimental campaign assessed the manufactured cold plate in contact with a commercial $20\,\mathrm{Ah}$ lithium-ion pouch cell. Figure~\ref{fig:experimental_assembly} shows the machined plate, pouch cell, thermal interface material (TIM), cover and back plates, spring fasteners, electrical connections, and surface thermocouples. A recirculating coolant loop supplied a water--glycol mixture (50:50 by volume) to the plate. The plate geometry and assembly configuration were held fixed during the flow-rate study; the electrochemical parameters of the commercial cell were not identified for direct, cell-specific model calibration.

Tests were organised into three nominal inlet-temperature conditions, T16, T20, and T24. These are test-condition labels, not the measured temperature of every run: the inlet temperatures at the reported thermal peaks are listed individually in Table~\ref{tab:experimental_absolute}. Within each condition, the nominal coolant mass-flow settings were $0.1$, $1.5$, $3.0$, and $4.5\,\mathrm{g\,s^{-1}}$; the measured mean flow rates are also given in Table~\ref{tab:experimental_absolute}. A single zero-flow experiment, conducted at a measured inlet temperature of $25.92\,^{\circ}\mathrm{C}$ at the reporting instant, provides the uncooled reference. It is not an independent zero-flow measurement for each of the three nominal inlet-temperature conditions. Accordingly, differences between runs reflect their recorded inlet temperatures as well as the imposed flow settings.

The cell was subjected to a time-varying charge--discharge programme using a programmable battery cycler. Surface temperatures were recorded at the instrumented thermocouple locations shown in Fig.~\ref{fig:experimental_assembly}; coolant inlet and outlet temperatures and mass flow were recorded concurrently. The spatial temperature maps in Figs.~\ref{fig:heatmap_peaks_troughs_T16} and~\ref{fig:heatmap_peaks_troughs_T24} were reconstructed by interpolation between measured locations. They are not direct, pixel-resolved infrared measurements. The independently measured pumping-power values used in the experimental performance map, Fig.~\ref{fig:experimental_pareto_pouch}, were associated with the corresponding flow settings; they are distinct from the pumping-loss objective evaluated in the numerical geometry optimisation.

For each run, the reporting instant was identified from the peak of the smoothed mean temperature across the instrumented cell locations. The maximum, mean, and minimum cell temperatures were evaluated simultaneously at that instant; the point-to-point range is $\Delta T_{\mathrm{range}}=T_{\max}-T_{\min}$. The inlet and outlet coolant temperatures reported in Table~\ref{tab:experimental_absolute} correspond to the same reporting instant. Temperature rises were formed by subtracting the measured inlet temperature of the respective run. The $\pm$ values in Table~\ref{tab:experimental_absolute} are standard deviations evaluated in a $100\,\mathrm{s}$ window centred on the selected peak. They describe within-run temporal variation, not sensor calibration uncertainty or between-run repeatability. Where the zero-flow reference appears in more than one plotted condition, the same recorded experiment was reused rather than counted as an additional replicate.

\section{Results}
\subsection{Evolutionary Optimization Results}
\begin{figure}
    \centering
    \includegraphics[width=0.9\textwidth]{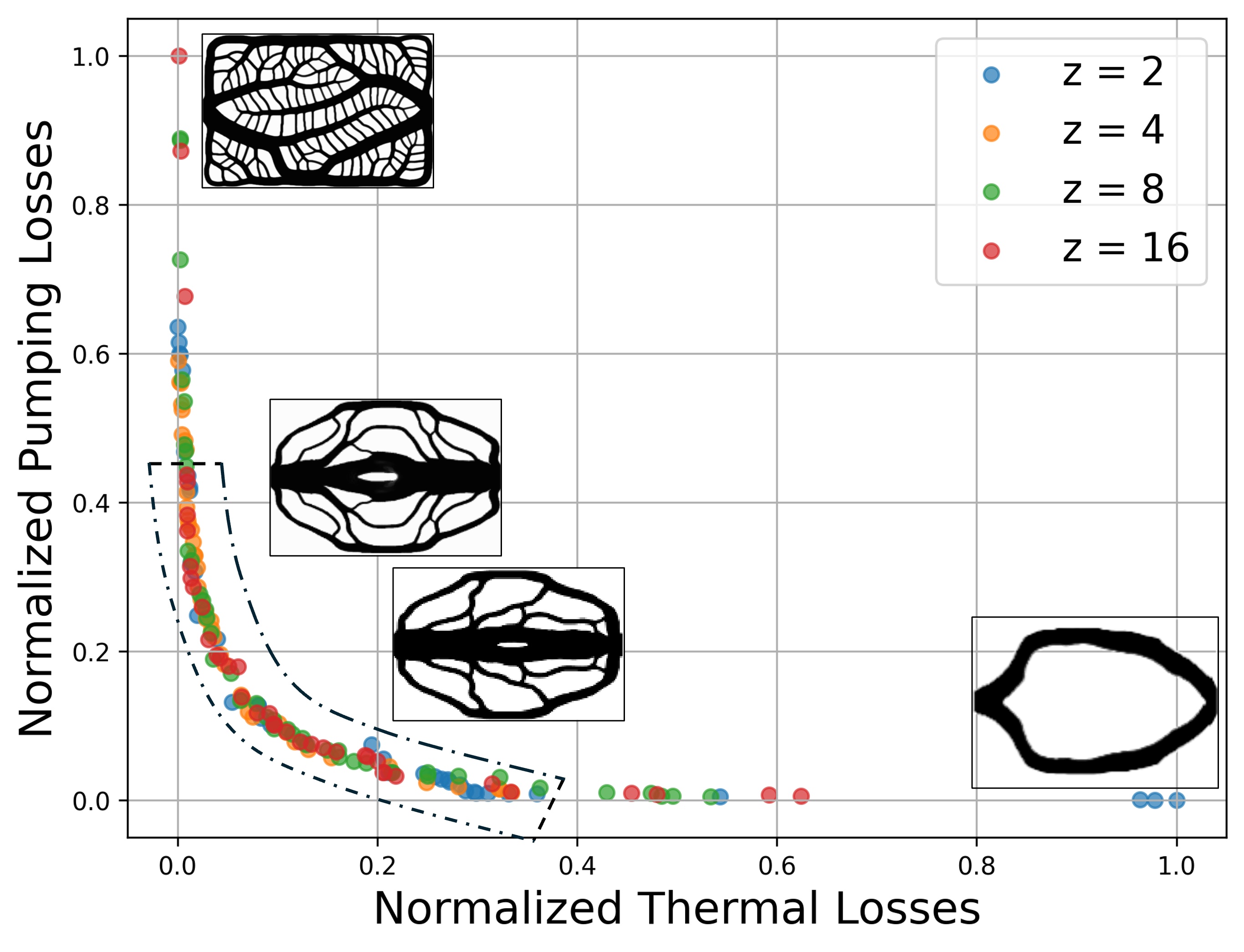}
    \caption{Pareto front illustrating the trade-offs between thermal losses and pumping losses, highlighting the diverse set of optimal solutions achieved through NSGA-II optimization. The normalized thermal and pumping losses are plotted for different values of latent space size ($z$), demonstrating the correlation between performance and freedom-to-morph. Insets show representative vascular networks corresponding to different trade-offs along the Pareto front. The dotted line highlights the candidates selected for further multi-objective enhancement using gradient-based optimization (Fig. \ref{fig:pareto3D}).}
    \label{fig:pareto}
\end{figure}

\begin{figure}
    \centering
    \includegraphics[width=0.9\textwidth]{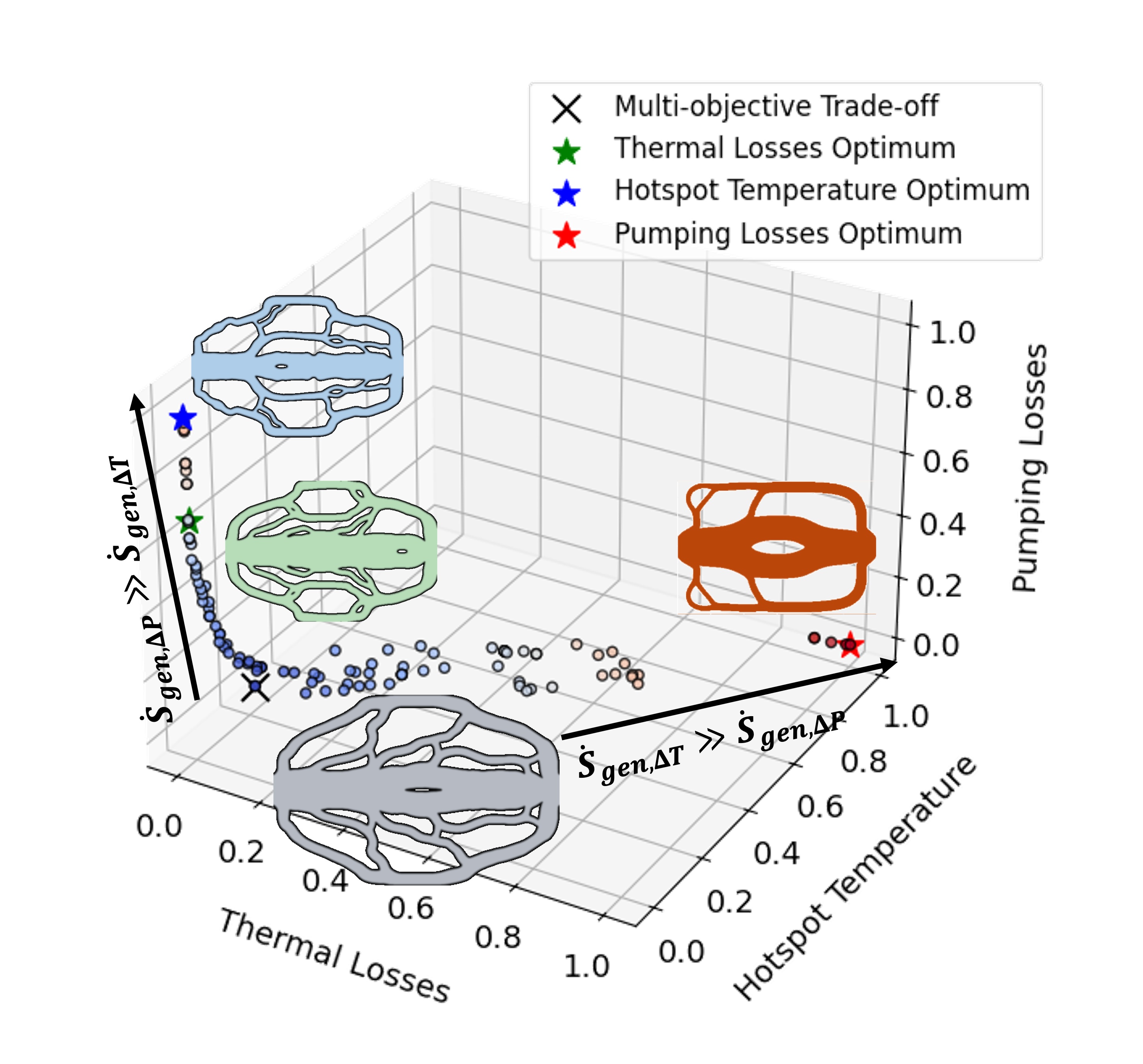}
    \caption{Expanded 3D Pareto front incorporating mean temperature, hotspot temperature, and pumping losses. This visualization extends the trade-offs shown in Fig. \ref{fig:pareto}, revealing the multi-objective landscape where the optimal solution balance all three competing objectives and minimizes entropy generation. The trade-off point (black cross) represents the knee of the Pareto front, while the single-objective optima (stars) highlight the single-objective optima. Insets illustrate representative vascular networks corresponding to different regions of the front, emphasizing the morphological adaptations that emerge to minimize thermal and pumping losses while controlling temperature hotspots.}
    \label{fig:pareto3D}
\end{figure}

The evolutionary optimization of the cold plate design was started using NSGA-II for a simplified physics model with uniform heat generation, with the objective of minimizing both thermal and pumping losses. 
The optimization process was further refined by iteratively updating the design variable $\gamma$ using the Globally Convergent Method of Moving Asymptotes (MMA) and evaluating performance through finite element analysis. 
Convergence was achieved when successive iterations exhibited negligible changes ($\leq 1 \times10^{-6}$) in thermal and pumping losses, ensuring globally optimized vascular networks.
The resulting Pareto fronts, shown in Figs. \ref{fig:pareto}-\ref{fig:pareto3D}, illustrate the trade-offs between competing performance metrics. 
Each point on the Pareto front represents a non-dominated solution, demonstrating the diverse set of optimal vascular network configurations that emerge from the optimization process.

The influence of key design variables, particularly the latent space size ($z$), is evident in the results in Fig. \ref{fig:pareto}. 
As $z$ increases, the optimization algorithm is granted greater freedom to explore complex morphologies, leading to a broader range of efficient solutions. 
This is reflected in the variation of thermal and pumping losses along the Pareto front, with some designs favouring minimal thermal resistance at the cost of higher pumping power, while others prioritize hydraulic efficiency with slightly elevated thermal losses.
Notably, the highly branched structures correspond to high latent space dimensionality ($z=8,16$), while the results streamlined for pumping losses correspond to low latent space dimensionality ($z=2$).

The insets in Figure \ref{fig:pareto} highlight representative vascular networks corresponding to different trade-offs along the front. 
These optimized designs exhibit varying levels of complexity, with some solutions featuring hierarchical branching structures that enhance heat dissipation while maintaining moderate pressure drop.

Building on the candidates discovered in Fig. \ref{fig:pareto}, the expanded Pareto front in Fig. \ref{fig:pareto3D} incorporates mean temperature, hotspot temperature, and pumping losses, providing a comprehensive view of the multi-objective optimization landscape.
In classical thermodynamic optimization using entropy generation minimization, Bejan established a trade-off between thermal resistance and pumping power by varying the mass flow rate \cite{Bejan1978}. 
Subsequently, the constructal law states that flow architectures evolve to provide progressively better access to the currents that move through them, leading to optimized configurations through geometric adaptation \cite{Bejan1997}. 
While the former approach optimizes performance by adjusting operating conditions, the latter achieves improvements by evolving system architecture. 
Consistent with constructal design principles, the present optimization fixes the mass flow rate while allowing the vascular network topology to adapt, enabling the emergence of configurations that balance thermal and hydraulic performance.

The results indicate distinct morphological adaptations depending on the optimization objective. 
Designs optimized for minimal pumping losses exhibit streamlined, high-conductance paths that minimize flow resistance. 
However, these configurations lead to localized thermal gradients (thermal losses) due to limited heat spreading. 
In contrast, networks optimized for thermal performance feature increased branching, which enhances heat dissipation. 
This improvement comes at the cost of higher pumping power (hydraulic losses) due to increased flow resistance.
The knee point of the Pareto front (black cross) represents the optimal balance among the three objectives, minimizing entropy generation. 
The insets highlight representative network configurations corresponding to different trade-offs along the front, demonstrating the morphological variations that emerge from the optimization process.
Most notably, the bucket shape of the Pareto frontier closely resembles the shape of the exergy loss variation with mass flow rate, discovered analytically in the context of thermoelectric power packaging (Fig. 8 in \cite{Bejan2014}).
Most notably, the shape of the Pareto frontier resembles the power loss variation with mass flow rate, as shown analytically in thermoelectric power packaging (Fig. 8 in \cite{Bejan2014}), highlighting the effectiveness of generative evolutionary optimization in automating constructal design.

\subsection{Numerical comparison with parametric cold plates}
\label{sec:parametric_benchmark}

\begin{figure}[!htbp]
    \centering
    \includegraphics[width=\textwidth]{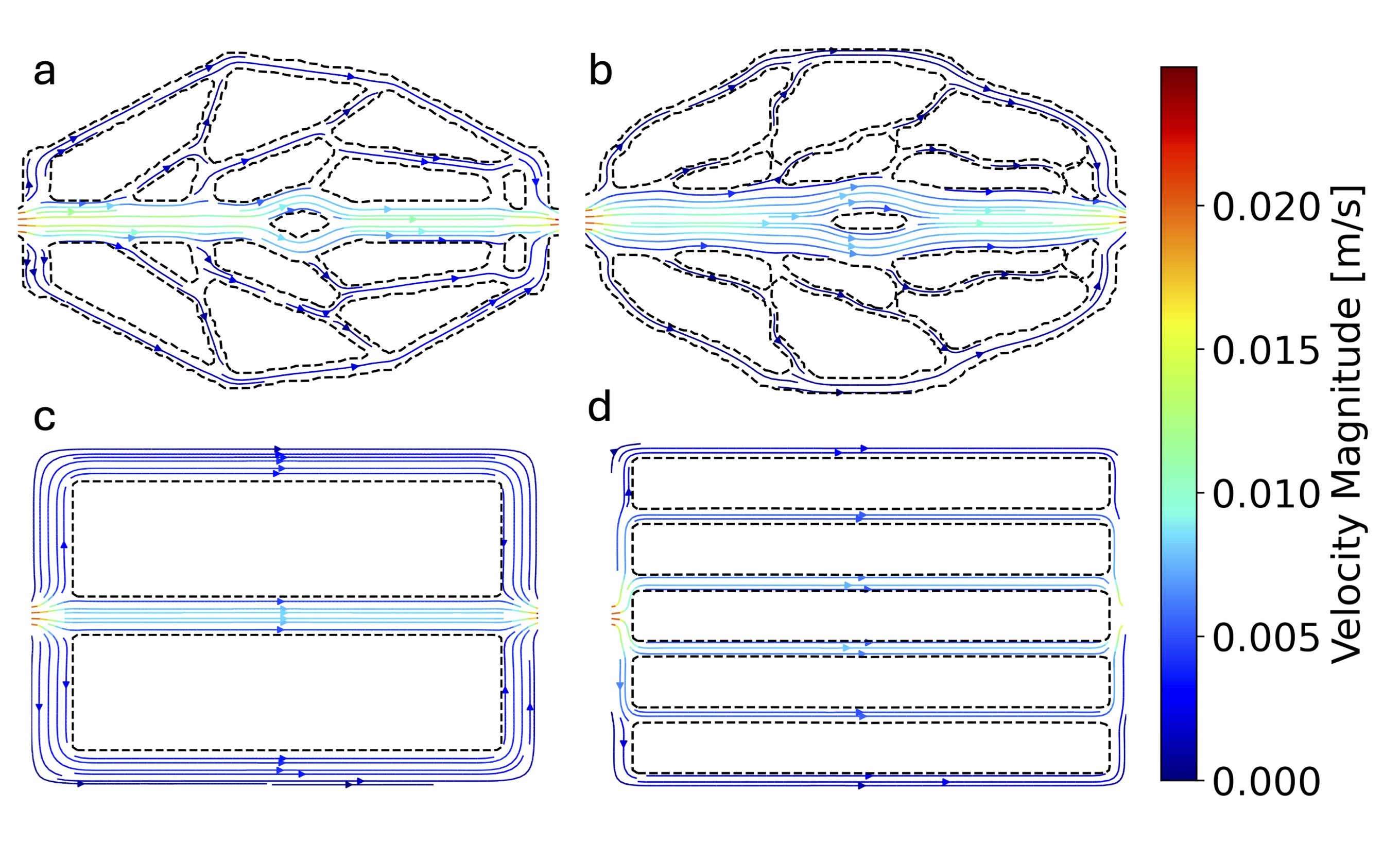}
    \caption{Computed coolant streamlines, coloured by velocity magnitude, for (a) the reconstructed network, (b) the organic evolutionary network, (c) the coarse parametric straight-channel plate, and (d) the fine parametric straight-channel plate. All four cases use the same inlet and outlet boundary conditions.}
    \label{fig:benchmark_streamlines}
\end{figure}

To assess the benefit of the evolved flow architecture independently of the subsequent pouch-cell experiment, the selected organic network was compared numerically with its straight-channel reconstruction and two parametric cold plates. The parametric cases use coarse and fine arrangements of uninterrupted channels between inlet and outlet headers, without tributaries within the active region. They provide manufacturable reference layouts for comparing the effects of channel distribution and branching; the four geometries are shown in Fig.~\ref{fig:benchmark_streamlines}. All cases were evaluated using the same coupled thermal--electrochemical model, prescribed coolant flow rate and inlet temperature, coolant and plate properties, battery loading, and nominal battery--plate interface condition. The comparisons in this subsection are simulation results, not separate experimental tests of the parametric plates.

\paragraph{Hydraulic response.}
The straight-channel layouts carry flow principally along continuous passages between the headers, whereas the organic and reconstructed networks provide additional paths into the active area. Figure~\ref{fig:benchmark_streamlines} illustrates the resulting differences in flow distribution: the reconstructed network retains the main access paths of the organic design but introduces angular junctions in place of curved branches. Under the nominal simulated condition reported here, the organic network required approximately half the pumping power of the fine parametric plate and approximately 12\% less than the coarse parametric plate. These comparisons refer to the \emph{geometry benchmark at fixed imposed flow}, not to the measured pumping-power changes between flow settings in the experimental operating map. The locations and magnitudes of local velocity maxima should not, by themselves, be interpreted as total pumping losses; the hydraulic comparison is based on the integrated objective used for the simulated cases.

\begin{figure}[!htbp]
    \centering
    \includegraphics[width=\textwidth]{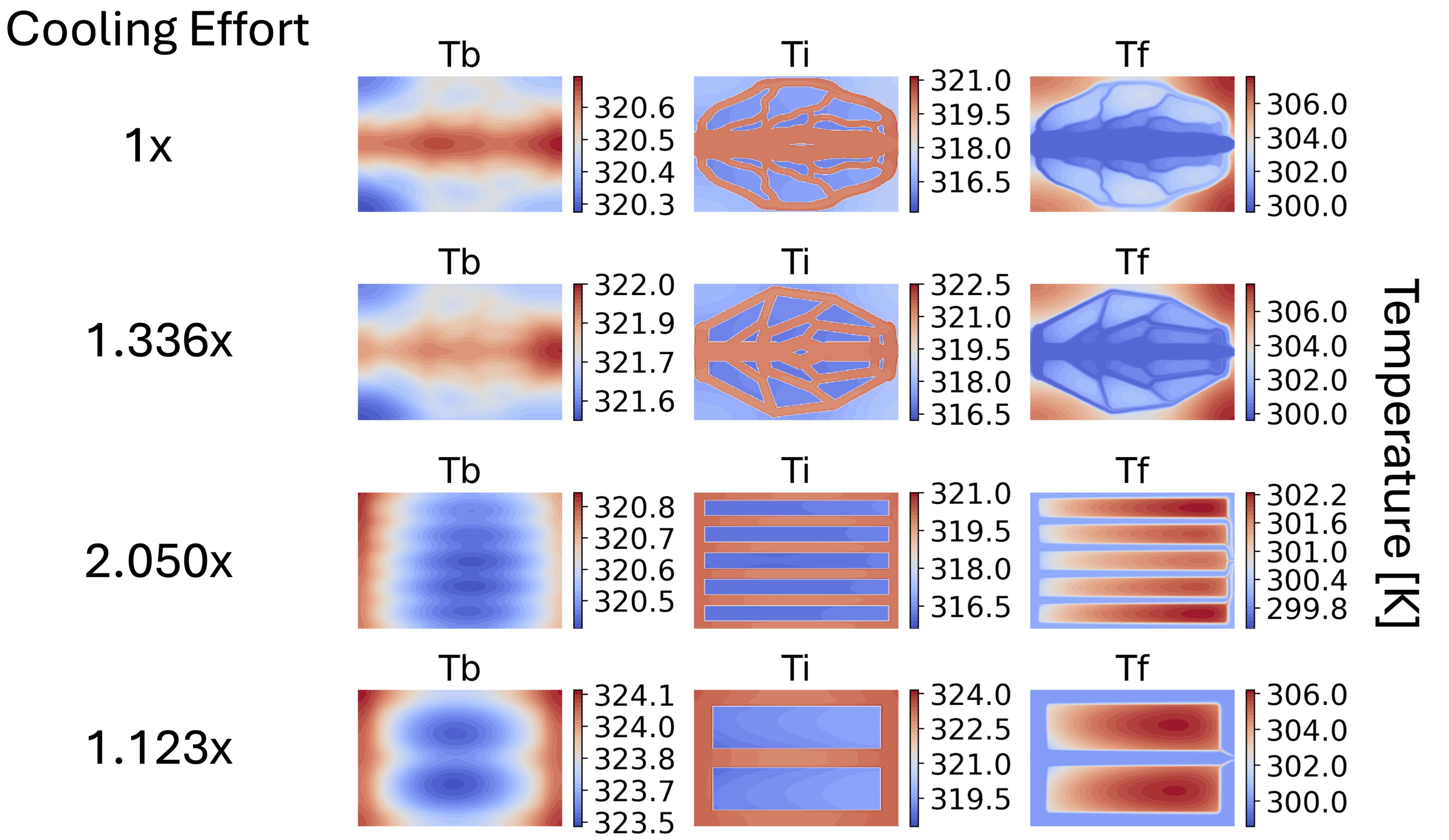}
    \caption{Simulated battery ($T_b$), interface ($T_i$), and cold-plate-domain ($T_f$) temperature fields at the selected peak load for the organic, reconstructed, fine parametric, and coarse parametric geometries (top to bottom). Colour scales differ between panels; the numerical limits, rather than colours alone, should be used for comparison. The left-hand cooling-effort annotations are relative to the organic case.}
    \label{fig:benchmark_temperatures}
\end{figure}

\paragraph{Battery-surface temperatures.}
The peak-load temperature fields in Fig.~\ref{fig:benchmark_temperatures} show why fluid-domain temperature alone is an incomplete measure of battery cooling. The parametric plates leave larger areas between passages dependent on in-plane heat spreading, while the distributed paths in the organic and reconstructed networks alter the temperature field across the battery footprint. The organic network gives the lowest mean-temperature and temperature-nonuniformity objectives among the illustrated cases at the selected peak-load condition. Its mean-temperature advantage is modest compared with the more pronounced separation in the temperature-gradient objective: the two metrics therefore describe different benefits of the architecture. The reconstructed design preserves much of the spatial access of the organic network despite straightening its channels, while its hydraulic performance differs from that of the original network. The reported comparison is specific to the common simulated load and boundary conditions; it does not establish an experimental percentage improvement over an untested conventional plate.

\begin{figure}[!htbp]
    \centering
    \includegraphics[width=\textwidth]{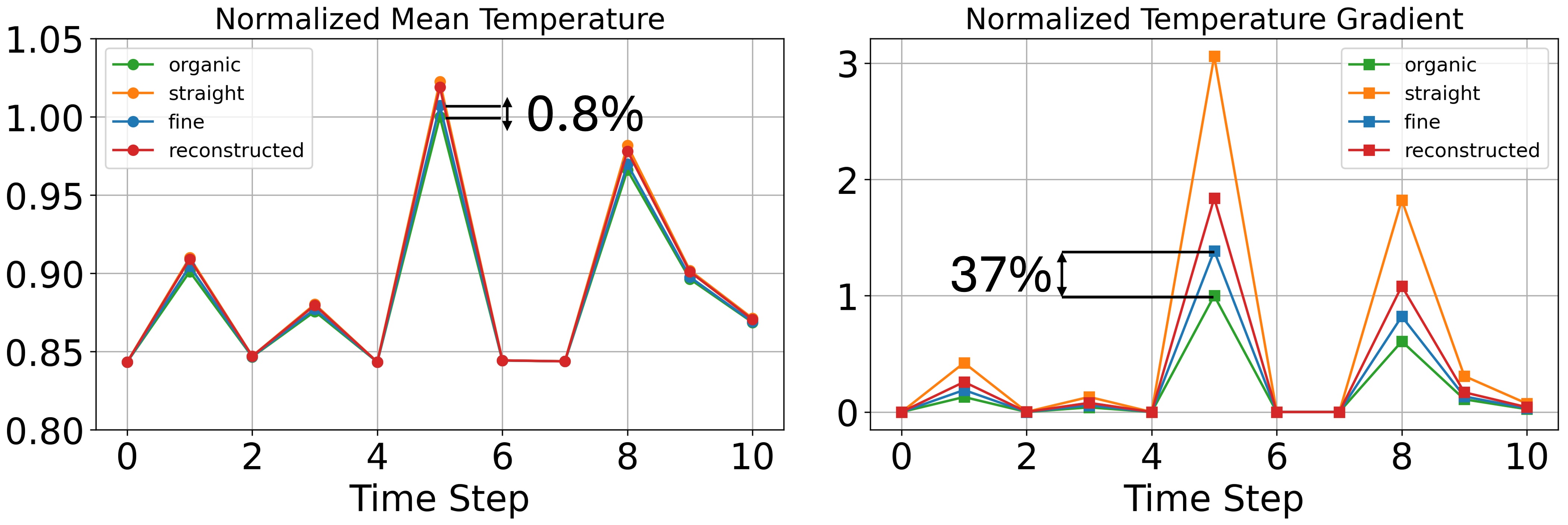}
    \caption{Normalized mean battery temperature (left) and temperature-gradient objective (right) at sampled times during the imposed simulated load sequence for the organic, reconstructed, fine parametric, and coarse parametric cold plates. All quantities are numerical results obtained under matched operating and interface conditions.}
    \label{fig:benchmark_transient}
\end{figure}

\paragraph{Response over the imposed load sequence.}
Figure~\ref{fig:benchmark_transient} compares the normalized mean battery temperature and temperature-gradient objective at the sampled times of the simulated operating sequence. The mean-temperature trajectories remain close across geometries, whereas the gradient objective separates strongly during the higher-load events. The organic design has the smallest gradient objective at the principal peaks shown, and the reconstructed design retains part of that improvement relative to the parametric cases. This result indicates that redistributing coolant access has a larger effect on transient spatial nonuniformity than on the domain-mean temperature for the conditions simulated. The figure reports sampled objective values; claims about recovery time or the maximum temperature between samples require the underlying time-resolved traces.

Taken together, the matched-geometry comparisons establish a numerical baseline for the evolutionary design: the organic and reconstructed flow networks can be assessed against straight-channel alternatives without conflating a change in plate geometry with a change in coolant operating condition. The pouch-cell tests reported in Section~\ref{sec:experimental_results} instead characterize one manufactured plate over a range of flow and inlet-temperature settings; they do not constitute an experimental comparison of the four geometries.

The numerical benchmark compares the selected organic topology with a straight-channel reconstruction and two parametric reference layouts under matched conditions. The organic network shown in Fig.~\ref{fig:benchmark_streamlines}(b) is the design taken forward from the simulated candidate set in Fig.~\ref{fig:pareto3D} for machining; Fig.~\ref{fig:benchmark_streamlines}(a) shows the reconstructed alternative used only in the numerical comparison. 

Figure~\ref{fig:experimental_assembly} links the organic design drawing to the as-built machined plate and its assembly with the pouch cell and thermal interface. The measurements in Section~\ref{sec:experimental_results} therefore characterize the manufactured organic network, not the straight-channel reconstruction. 

\subsection{Thermal Interface Influence}
\begin{figure}
    \centering
    \includegraphics[width=\textwidth]{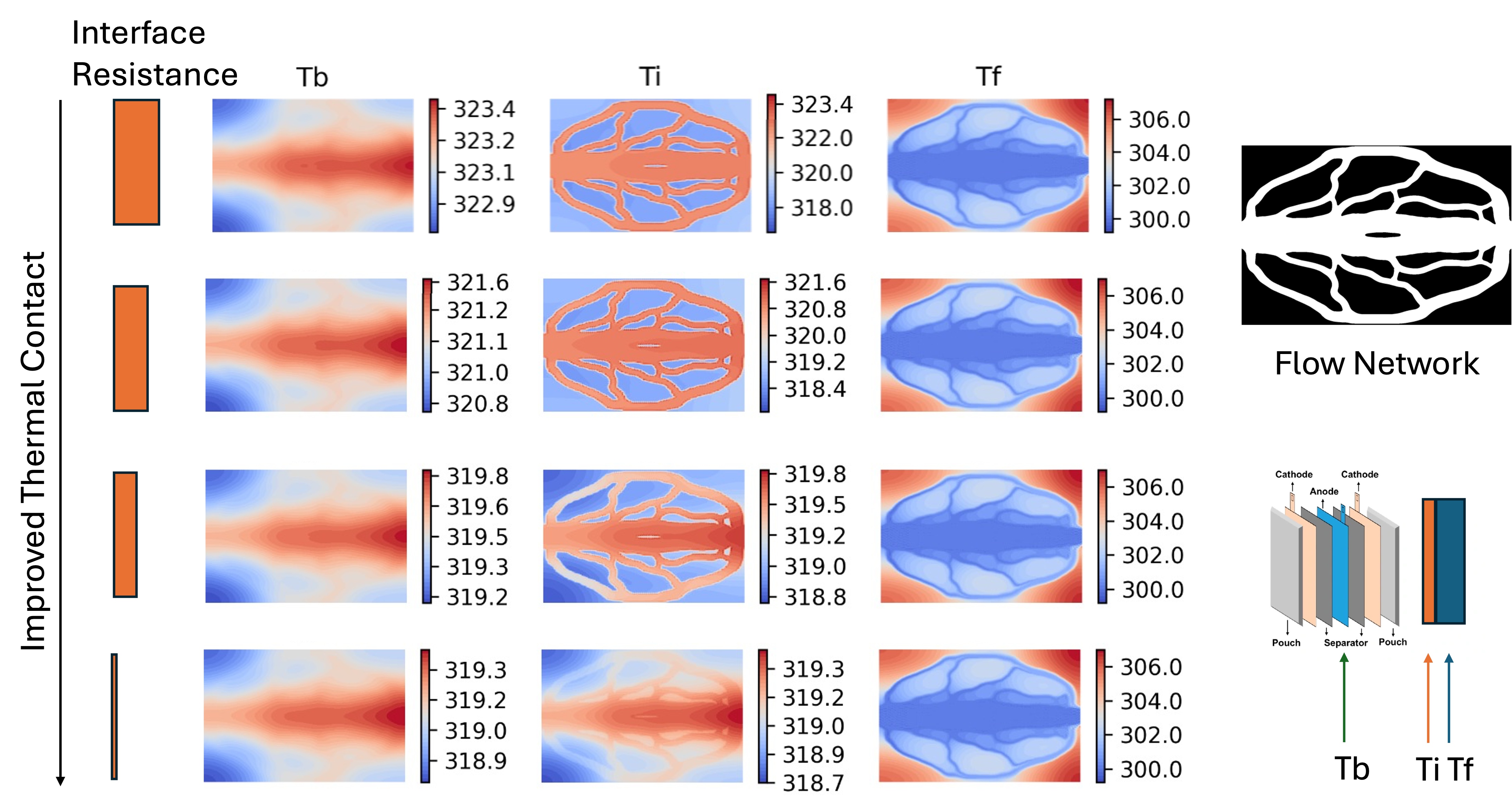}
    \caption{Temperature distributions for varying thermal interface resistances between the cold plate and battery. The left column shows the temperature at the core of the battery ($T_b$), the middle column illustrates the battery surface temperature ($T_i$), and the right column presents the temperature inside the cold plate ($T_f$). As interface resistance increases, peak battery temperatures rise, and thermal gradients intensify, especially above fluid channels where heat dissipation is limited by lower thermal conductance.}
    \label{fig:interface_temp}
\end{figure}

This analysis isolates the impact of thermal resistance between the cold plate and battery while keeping the flow network fixed. 
Interface resistance arises from roughness on the battery or cold plate surface, contact pressure, and thermal interface material (TIM). 
Lower resistance improves cooling but may require advanced TIMs, higher contact pressure, or surface treatments --introducing trade-offs in cost, complexity, and mechanical stress. 
By systematically adjusting this parameter, we examine how thermal resistance influences the temperature distributions and overall cooling effectiveness in all of the three layers modelled.
Figure \ref{fig:interface_temp} illustrates these effects, showing temperature distributions for varying interface resistances.

At lower resistance values, heat is efficiently conducted away, resulting in a more uniform thermal profile, with temperature variations below $\textbf{1K}$. 
As thermal resistance increases, heat dissipation becomes less efficient, leading to elevated peak temperatures and more pronounced thermal gradients at the surface of the battery. 
Reduced thermal coupling between the cold plate and the battery restricts heat flow, leading to localized overheating, particularly in areas directly above the fluid channels where heat dissipation is also limited by the  conduction bottleneck.
This phenomenon is evident in Fig. \ref{fig:interface_temp}, where higher interface resistance correlates with higher hotspot temperature in the middle column.

Most importantly, temperature distribution in the fluid layer, which was commonly used as a suitable metric for objective functions in gradient-based optimization \cite{Chen2022, Alexandersen2020}, proves to be insensitive to the change in thermal resistance. 
This renders it ineffective for the purpose of thermal design, as it does not reflect the impact of interface resistance on overall thermal performance.

As thermal interface resistance decreases, more cooling power becomes available due to the reduction in temperature-driven losses. 
This enhanced heat dissipation capacity allows for increased flexibility in the design of the flow network. 
With improved thermal contact, the flow network can afford more branching (increased svelteness \cite{IgnutaBotas2025}), as the reduced thermal resistance facilitates more effective heat transfer from the battery to the cold plate. 
Consequently, the network design can incorporate more fluid channels and finer geometries, enhancing the overall cooling effectiveness.

Conversely, at high thermal interface resistance, cooling performance becomes limited by conductive heat transfer. 
In this case, the flow network evolves to thicker configurations, as the returns from branching are more easily outweighed by pumping losses. 
In addition, the impact of the cooling network on the battery temperature is reduced, as the "blanket" formed by the interface limits the communication with the cold plate.
Moreover, thicker channels with higher cooling inertia are needed to handle the higher temperatures at the battery surface.

\begin{figure}
    \centering
    \includegraphics[width=\textwidth]{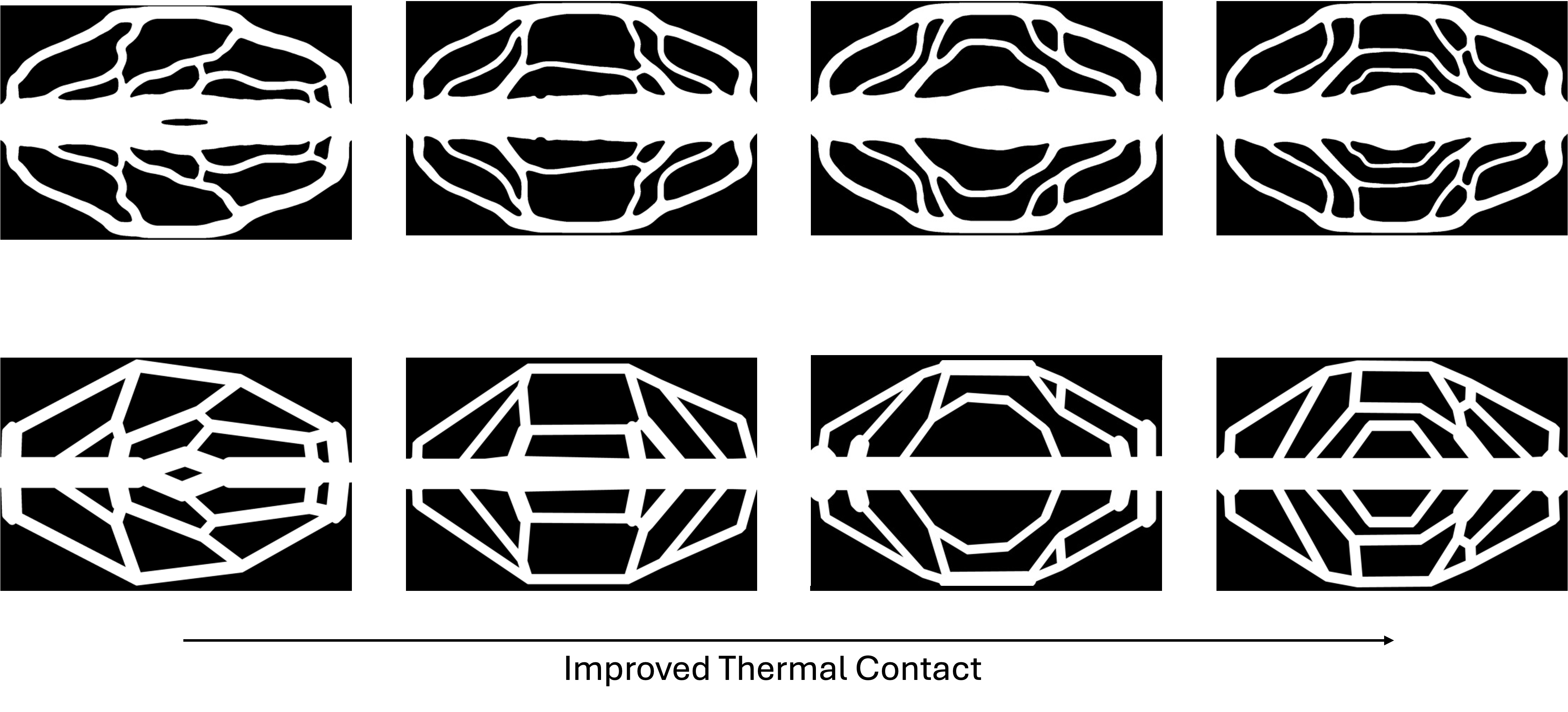}
    \caption{Optimized flow network designs for varying thermal interface resistance. From left to right, thermal contact improves, resulting in increased branching and finer channel structures in both \textit{organic} and straight vascular flow networks. This progression highlights how, when balancing thermal and fluidic losses, enhanced thermal coupling enables more svelte configurations, whereas higher interface resistance favours thicker channel designs.}
    \label{fig:interface}
\end{figure}

The shift in the optimal network morphology observable in Fig. \ref{fig:interface} emphasizes the interplay between flow geometry and thermal interface properties, showing how both thermal and fluidic factors shape the design. 
Capturing these complex interactions requires high-fidelity thermal-electrochemical modelling, while evolutionary design is also required for efficient exploration and high freedom-to-morph, with minimal computational cost.

\subsection{Performance Analysis of the Cold-Plate Design}
\label{sec:experimental_results}
\begin{figure}[!htbp]
\centering\includegraphics[width=\textwidth]{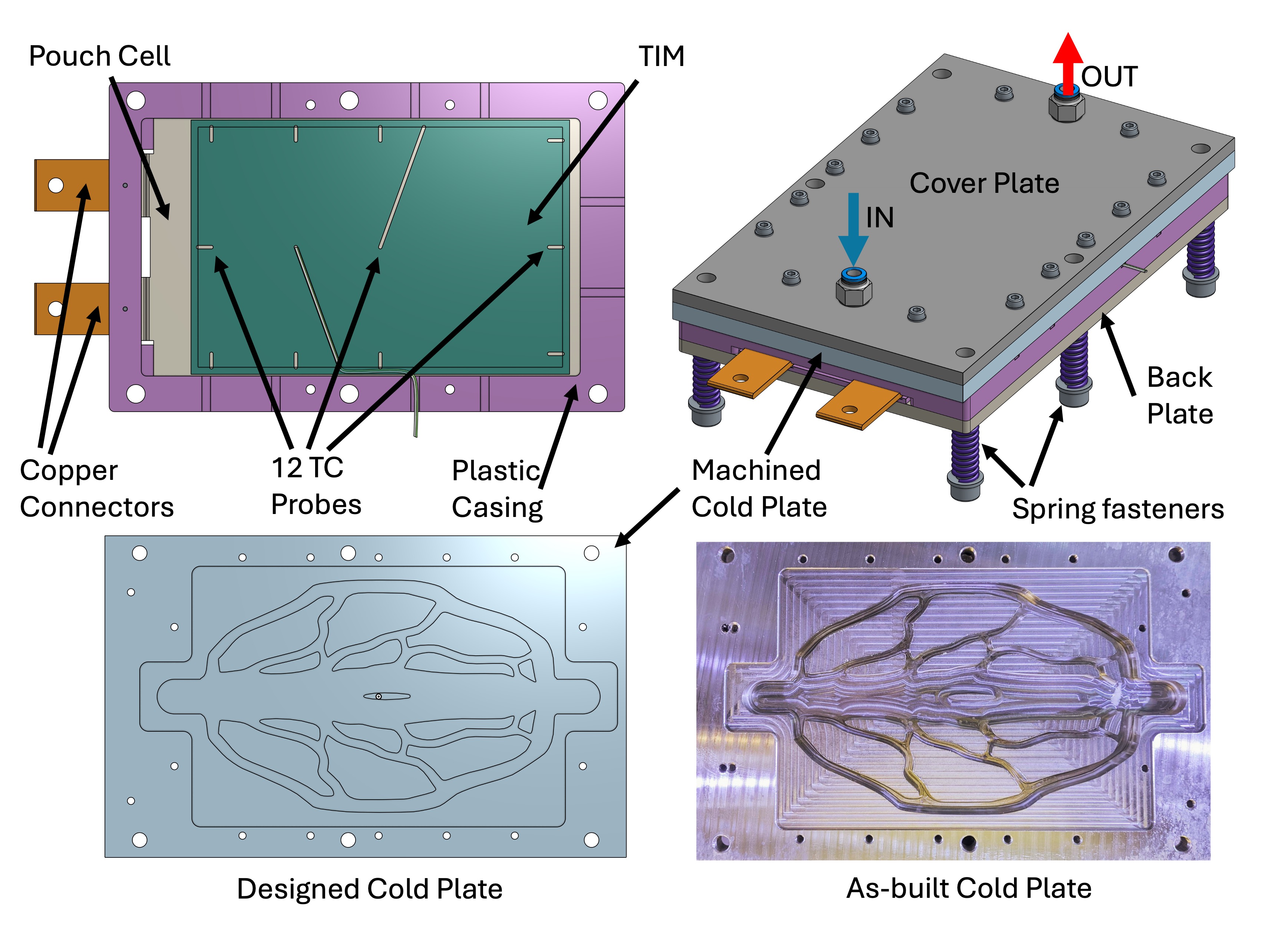}
\caption{Experimental pouch-cell cooling assembly, showing the thermal interface, machined cold plate, cover and back plates, spring fasteners, thermocouples, and coolant connections.}
\label{fig:experimental_assembly}
\end{figure}

The manufactured cold plate and pouch-cell test assembly are shown in Fig.~\ref{fig:experimental_assembly}. The tested inputs were nominal coolant inlet-temperature conditions T16, T20, and T24 and imposed mass-flow settings of $0.1$, $1.5$, $3.0$, and $4.5\,\mathrm{g\,s^{-1}}$, with one shared uncooled reference. The measured inlet temperatures and mass flows are given in Table~\ref{tab:experimental_absolute}; the condition labels denote nominal settings, not identical measured inlet temperatures. The plate geometry remained fixed. The simulated \emph{geometry} Pareto set in Fig.~\ref{fig:pareto3D} is distinct from the experimental \emph{operating-condition} map discussed below. Because the commercial pouch cell was not fully parameterized, these tests assess system response, not one-to-one agreement with simulated absolute battery temperature.

\subsubsection{Measured response to coolant flow}
\begin{figure}[!htbp]
\centering\includegraphics[width=\textwidth]{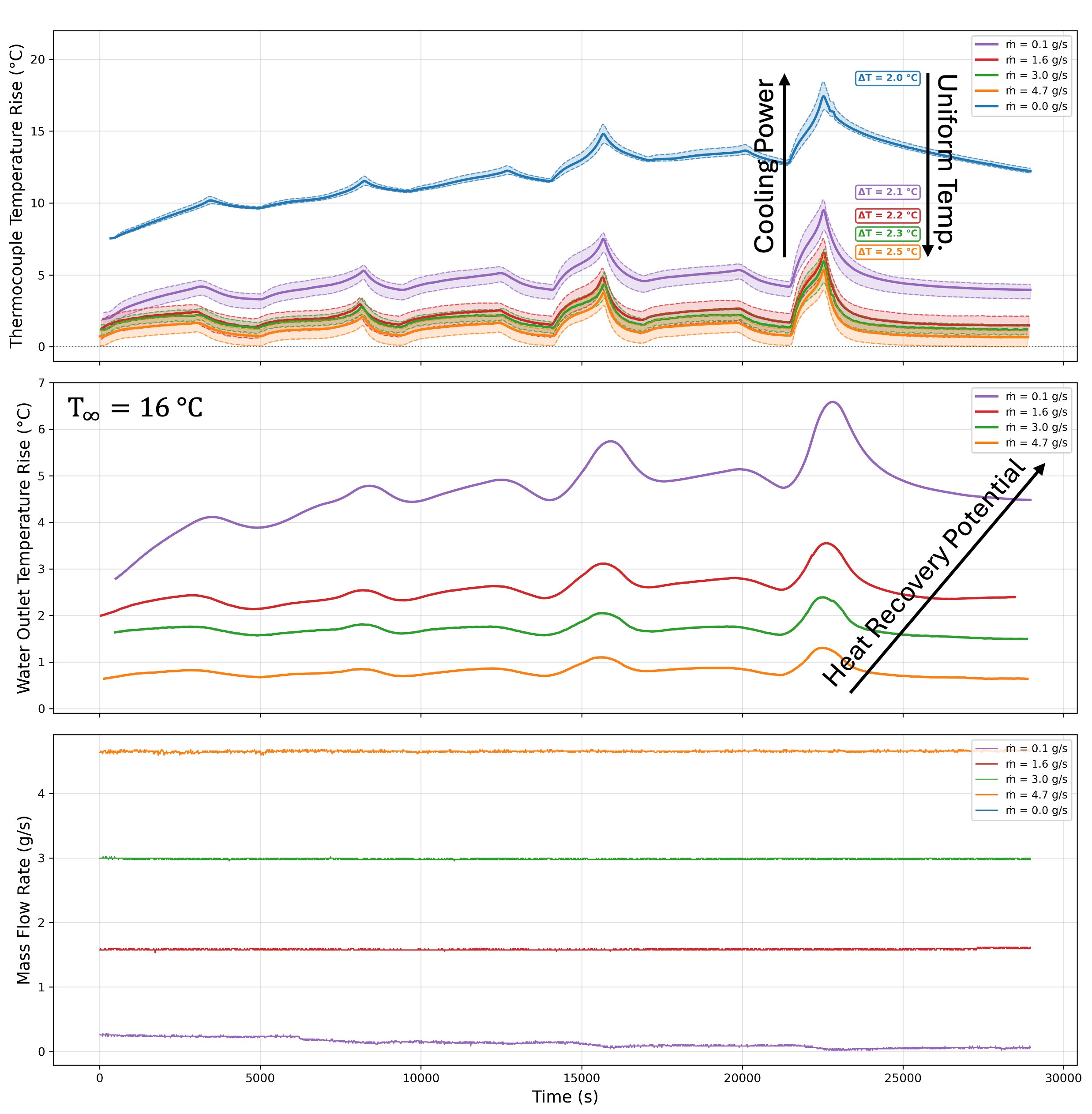}
\caption{Measured thermocouple temperature rise, outlet-coolant temperature rise, and mass-flow rate during the illustrated operating cycle.}
\label{fig:massflow_effects_pouch}
\end{figure}

Figure~\ref{fig:massflow_effects_pouch} shows thermocouple, outlet-water-temperature, and mass-flow histories. Mass flow remained approximately steady within each test, whereas the imposed battery load produced repeated thermal events. Increasing flow generally decreased surface temperature rise, but the benefit was not monotonic in every intermediate comparison. For T16, the measured maximum-temperature rises relative to the inlet were $7.62$, $7.81$, $6.92$, and $6.65\,^{\circ}\mathrm{C}$ at nominal flows of $0.1$, $1.5$, $3.0$, and $4.5\,\mathrm{g\,s^{-1}}$. The inlet temperature also varied between tests, so these comparisons do not isolate a purely causal flow effect. Coolant outlet temperature alone does not quantify heat-removal rate: a heat balance requires synchronized inlet and outlet temperatures, mass flow, coolant heat capacity, and an allowance for transient thermal storage.

\subsubsection{Transient thermal behaviour}
\begin{figure}[!htbp]
\centering\includegraphics[width=\textwidth]{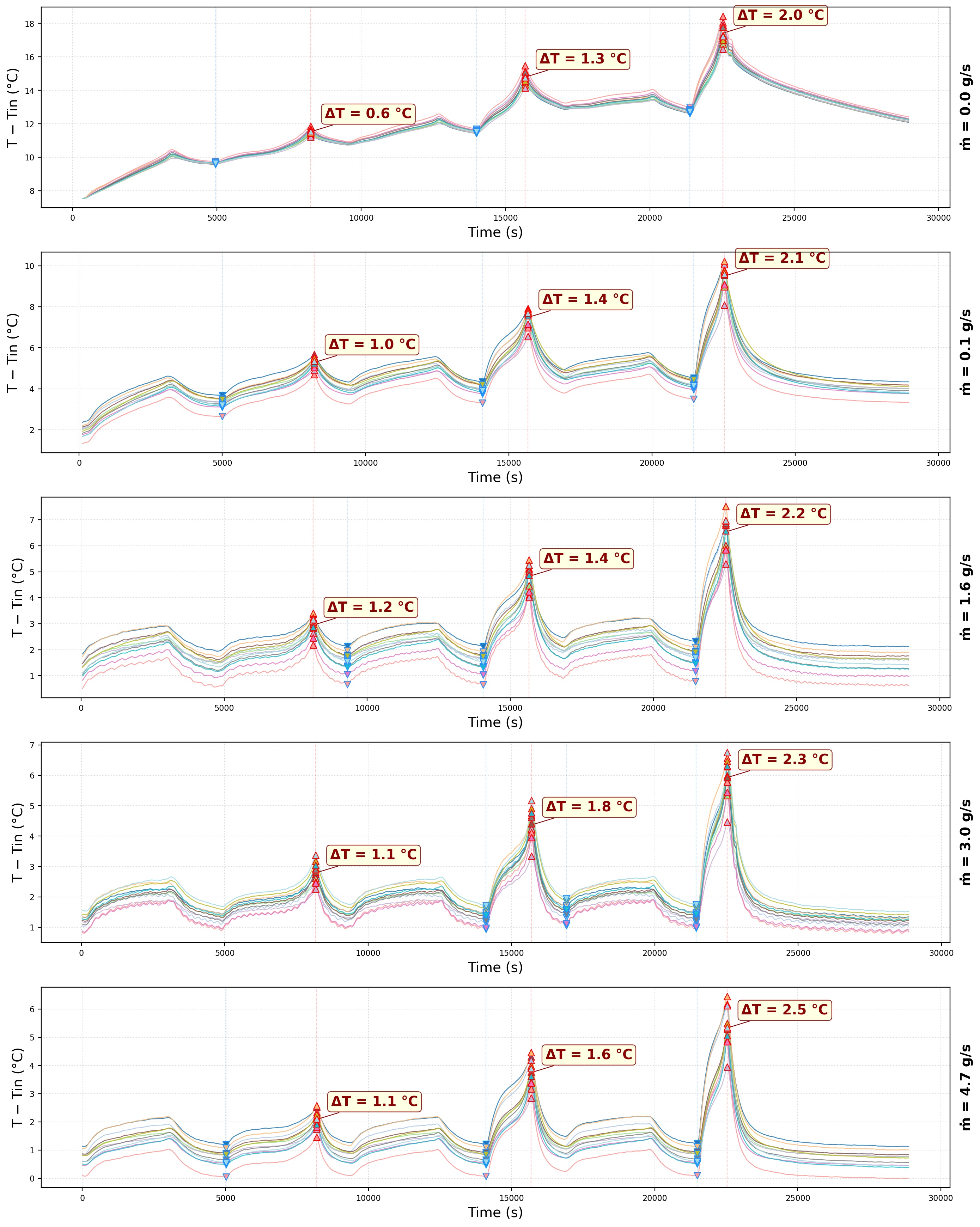}
\caption{Time-resolved temperatures at the instrumented locations for T16; marked events are used in Fig.~\ref{fig:heatmap_peaks_troughs_T16}.}
\label{fig:timeseries_peaks_troughs_T16}
\end{figure}

\begin{figure}[!htbp]
\centering\includegraphics[width=\textwidth]{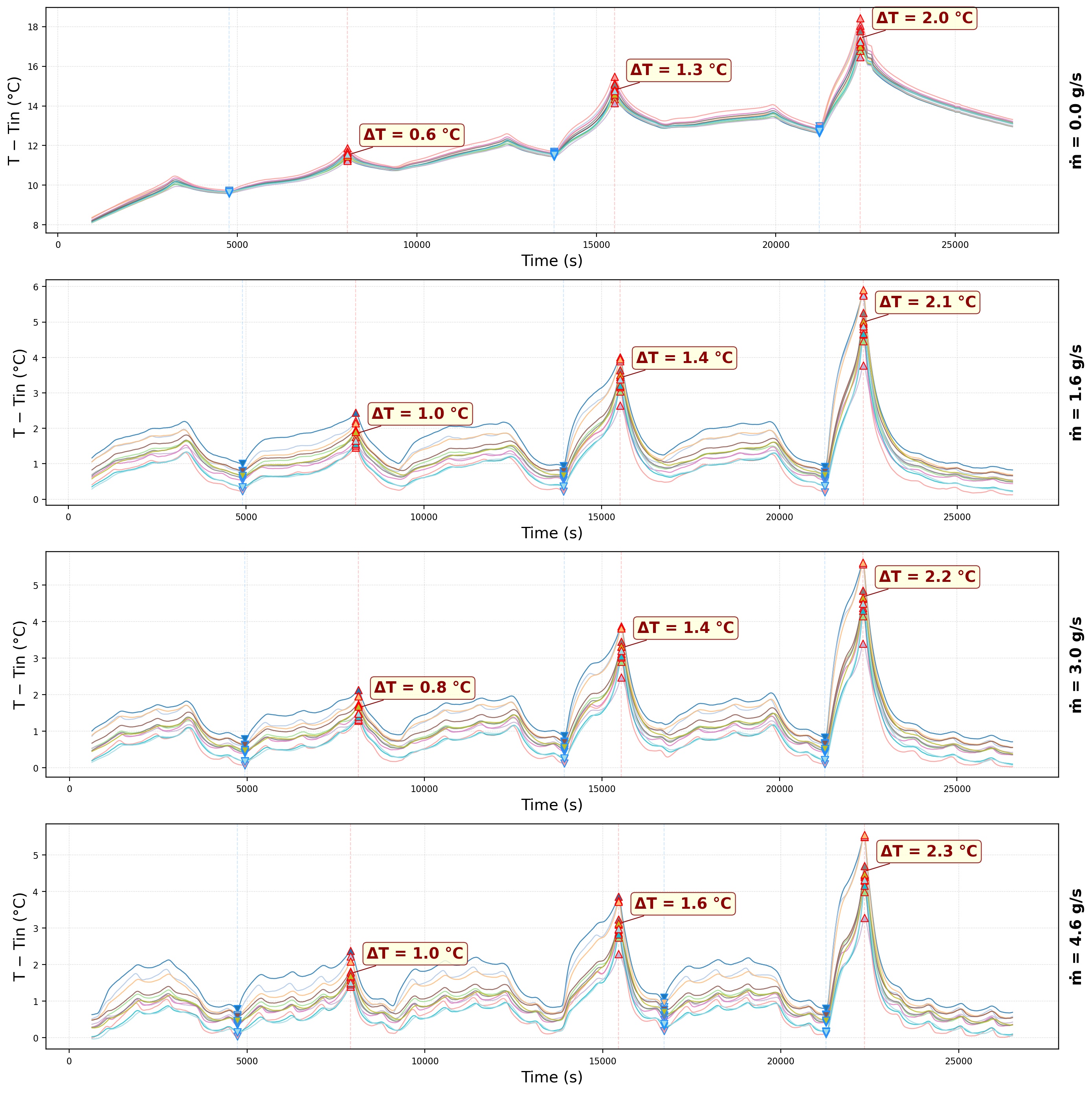}
\caption{Time-resolved temperatures at the instrumented locations for T24; marked events are used in Fig.~\ref{fig:heatmap_peaks_troughs_T24}.}
\label{fig:timeseries_peaks_troughs_T24}
\end{figure}

The time histories in Figs.~\ref{fig:timeseries_peaks_troughs_T16} and~\ref{fig:timeseries_peaks_troughs_T24} show heating during the repeated high-power events and subsequent cooling and redistribution. Separation between sensor traces changes during each event, indicating that a peak temperature alone does not characterize the transient spatial response. The marked peaks and troughs identify the events used to reconstruct the maps below.


For Table~\ref{tab:experimental_absolute}, the reporting instant was selected from the peak of the smoothed mean sensor temperature. Absolute temperatures and point-to-point range refer to that instant; the accompanying standard deviations were calculated over the surrounding $100$-s window. These are within-window variations, not between-test repeatability or measurement accuracy. The zero-flow observation was acquired once, at $25.92\,^{\circ}\mathrm{C}$ inlet temperature; its duplication across T16, T20, and T24 in the plotting data does not constitute three separate tests.

\begin{table}[!htbp]
\centering
\caption{Experimental quantities at the identified mean-temperature peak. The $\pm$ values are standard deviations within a $100$-s window, not independent-test repeatability.}
\label{tab:experimental_absolute}
\scriptsize\resizebox{\textwidth}{!}{%
\begin{tabular}{llrrrrrr}
Condition & $\dot m_{\mathrm{set}}$ & $\dot m_{\mathrm{meas}}$ & $T_{\mathrm{in}}$ & $T_{\max}$ & $T_{\mathrm{mean}}$ & $\Delta T_{\mathrm{range}}$ & $T_{\mathrm{out}}$ \\
 & (g s$^{-1}$) & (g s$^{-1}$) & ($^{\circ}$C) & ($^{\circ}$C) & ($^{\circ}$C) & ($^{\circ}$C) & ($^{\circ}$C) \\
\hline
Reference & 0.0 & 0.00 & 25.92 & $34.49\pm0.09$ & $33.49\pm0.06$ & $1.98\pm0.11$ & $25.90\pm0.01$ \\
\hline
T16 & 0.1 & 0.13 & 17.84 & $25.46\pm0.19$ & $24.75\pm0.16$ & $2.22\pm0.14$ & $21.66\pm0.11$ \\
T16 & 1.5 & 1.58 & 15.13 & $22.94\pm0.18$ & $22.02\pm0.18$ & $2.29\pm0.11$ & $18.95\pm0.04$ \\
T16 & 3.0 & 2.98 & 15.09 & $22.01\pm0.20$ & $21.13\pm0.14$ & $2.43\pm0.13$ & $17.63\pm0.07$ \\
T16 & 4.5 & 4.65 & 15.74 & $22.39\pm0.17$ & $21.29\pm0.16$ & $2.58\pm0.11$ & $17.18\pm0.05$ \\
\hline
T20 & 0.1 & 0.13 & 19.58 & $27.03\pm0.18$ & $26.07\pm0.15$ & $2.52\pm0.16$ & $22.98\pm0.11$ \\
T20 & 1.5 & 1.56 & 18.99 & $25.78\pm0.15$ & $24.81\pm0.15$ & $2.52\pm0.13$ & $21.47\pm0.07$ \\
T20 & 3.0 & 3.03 & 18.87 & $25.63\pm0.19$ & $24.47\pm0.16$ & $2.60\pm0.14$ & $20.62\pm0.06$ \\
T20 & 4.5 & 4.63 & 17.73 & $24.22\pm0.15$ & $23.10\pm0.15$ & $2.59\pm0.10$ & $18.99\pm0.04$ \\
\hline
T24 & 0.1 & 0.18 & 22.59 & $29.36\pm0.19$ & $28.48\pm0.16$ & $2.41\pm0.16$ & $25.37\pm0.12$ \\
T24 & 1.5 & 1.64 & 22.21 & $28.39\pm0.19$ & $27.50\pm0.16$ & $2.22\pm0.14$ & $24.11\pm0.07$ \\
T24 & 3.0 & 2.95 & 22.12 & $27.98\pm0.15$ & $27.06\pm0.14$ & $2.32\pm0.11$ & $23.55\pm0.05$ \\
T24 & 4.5 & 4.63 & 22.27 & $27.83\pm0.19$ & $26.84\pm0.15$ & $2.36\pm0.14$ & $23.23\pm0.06$ \\
\hline
\end{tabular}%
}
\end{table}

For T24, the maximum-temperature rise relative to the measured inlet decreased from $6.77$ to $5.56\,^{\circ}\mathrm{C}$ between nominal flows of $0.1$ and $4.5\,\mathrm{g\,s^{-1}}$. From $3.0$ to $4.5\,\mathrm{g\,s^{-1}}$, the reduction was only $0.31\,^{\circ}\mathrm{C}$; for T20 it was $0.27\,^{\circ}\mathrm{C}$. These observed differences are not measurements at perfectly identical inlet temperature and heat load. Meanwhile the point-to-point range remained approximately $2.0$--$2.6\,^{\circ}\mathrm{C}$ and did not decrease monotonically with flow. Lower temperature elevation and greater spatial uniformity must therefore be assessed separately.

\subsubsection{Spatial temperature distributions}
\label{sec:spatial_temperature_distributions}
\begin{figure}[!htbp]
\centering\includegraphics[width=0.75\textwidth]{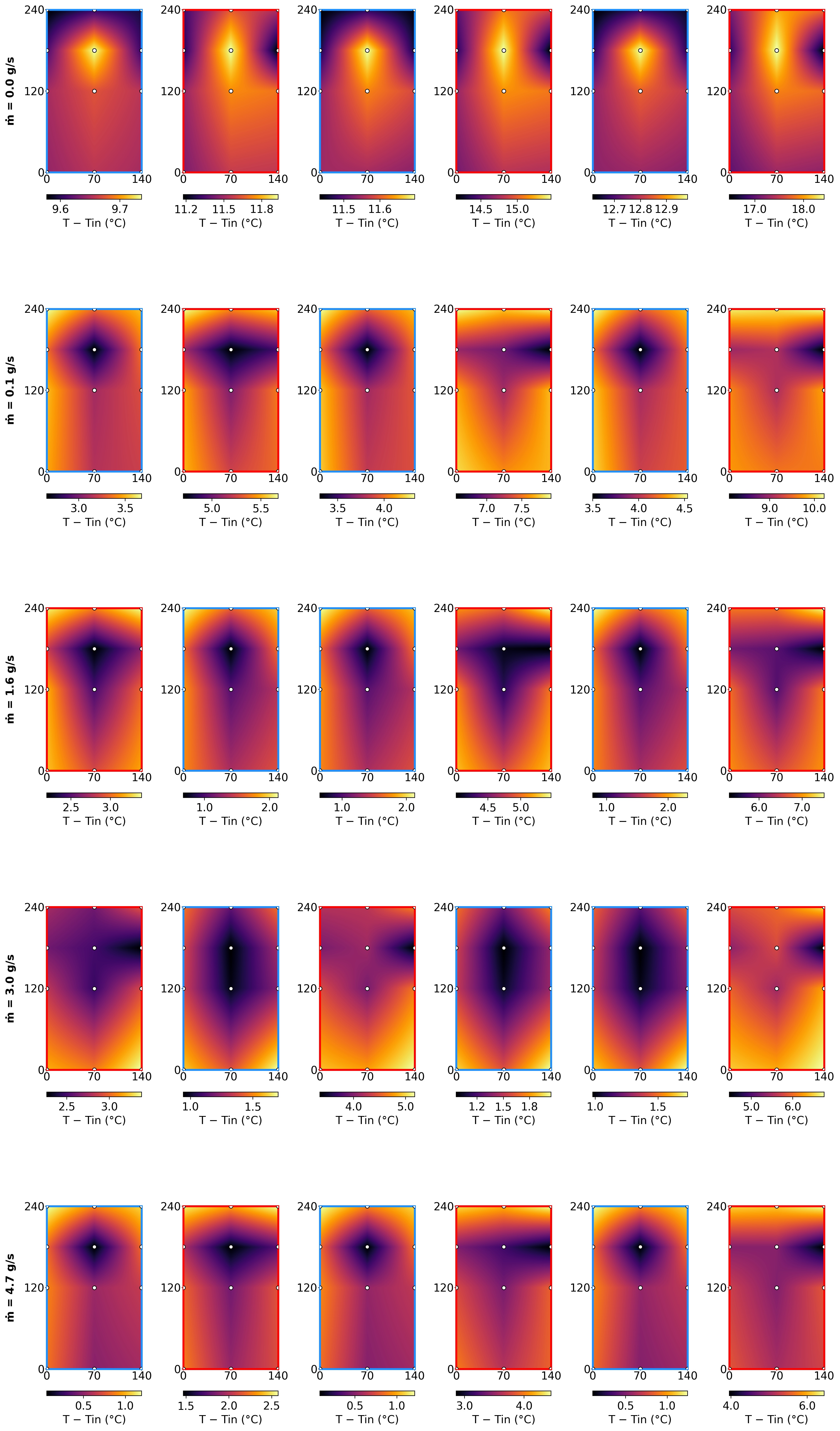}
\caption{Interpolated sensor-based temperature-rise maps at selected peaks and troughs for T16. Colour-bar limits differ across panels.}
\label{fig:heatmap_peaks_troughs_T16}
\end{figure}

\begin{figure}[!htbp]
\centering\includegraphics[width=0.75\textwidth]{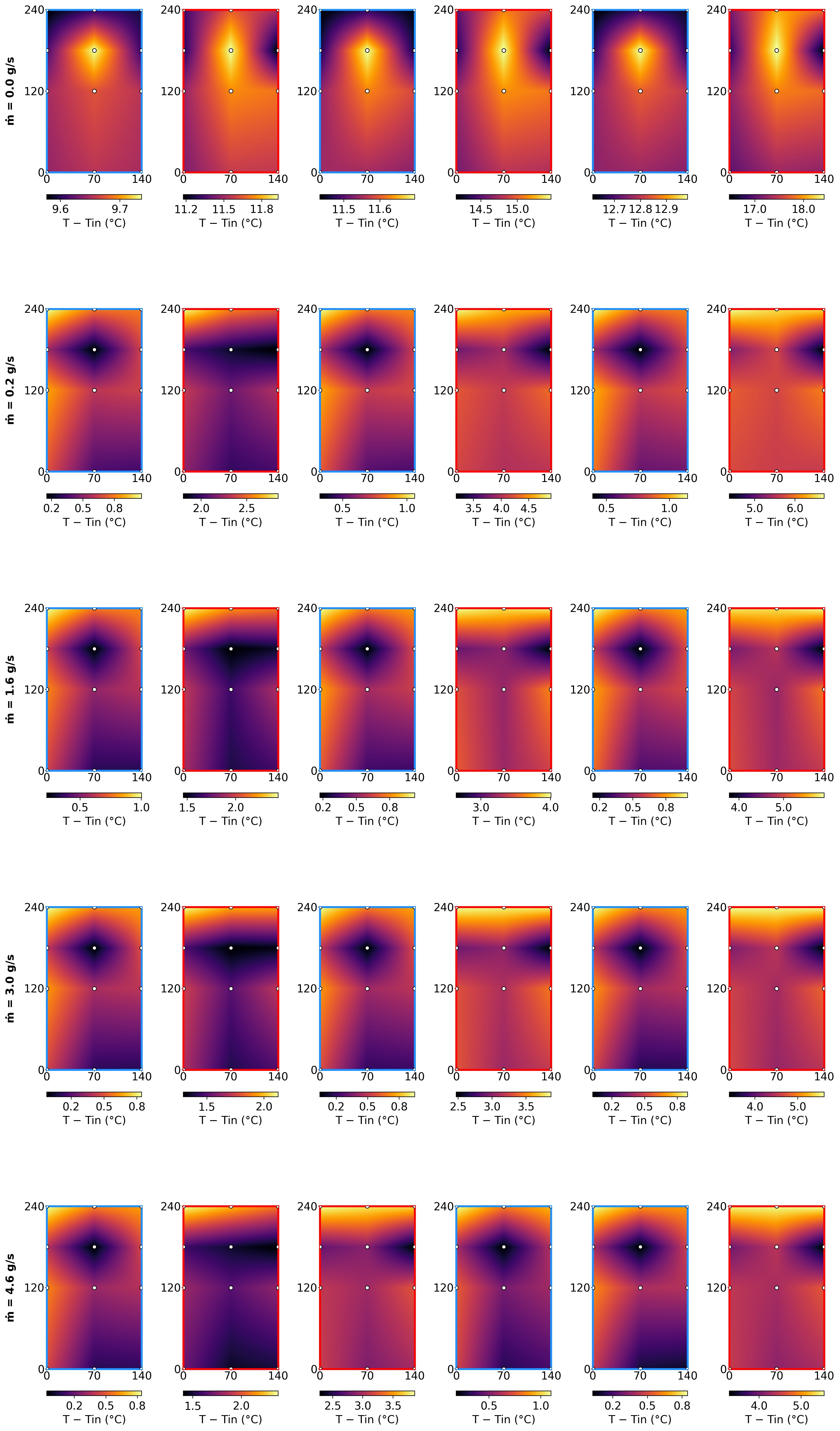}
\caption{Interpolated sensor-based temperature-rise maps at selected peaks and troughs for T24. Compare numerical colour-bar values, not colour alone.}
\label{fig:heatmap_peaks_troughs_T24}
\end{figure}

The T16 and T24 maps in Figs.~\ref{fig:heatmap_peaks_troughs_T16} and~\ref{fig:heatmap_peaks_troughs_T24} interpolate between instrumented locations; they are not direct full-field measurements. Panel colour scales differ, so identical colours in different maps do not imply identical temperatures. The marked numerical scales and time traces should be used to compare cases. At zero flow the selected events display a broad region of elevated temperature. Under forced flow the reconstructed field retains marked central-to-peripheral variation, even when surface-to-inlet temperature rise falls. The apparent location of the maximum is not identical across events and flow settings; this observation alone cannot distinguish transient heat generation, local cooling, interpolation effects, or interface variations.

For T16, higher-flow conditions generally have lower measured maximum and mean temperature rise, but the instantaneous point-to-point temperature range remains finite. Regions near and away from the main flow corridor can respond differently, although sensor interpolation does not establish a one-to-one correspondence with individual subsurface channels. T24 shows the same broad pattern: the mean and maximum rises decrease as flow increases without a monotonic reduction in the measured range.

Here $T_{\max}-T_{\mathrm{in}}$ describes the highest observed elevation relative to the inlet, whereas $\Delta T_{\mathrm{range}}=T_{\max}-T_{\min}$ is the simultaneous range across sensors. The two are distinct optimization criteria. The experimental maps demonstrate spatially nonuniform cooling but do not validate predicted cell-specific heat-generation patterns, because the commercial cell parameters are unknown.

\subsubsection{Optimal operation: experimental performance map}
\label{sec:optimal_operation}

\begin{figure}[!htbp]
    \centering
    \includegraphics[width=\textwidth]{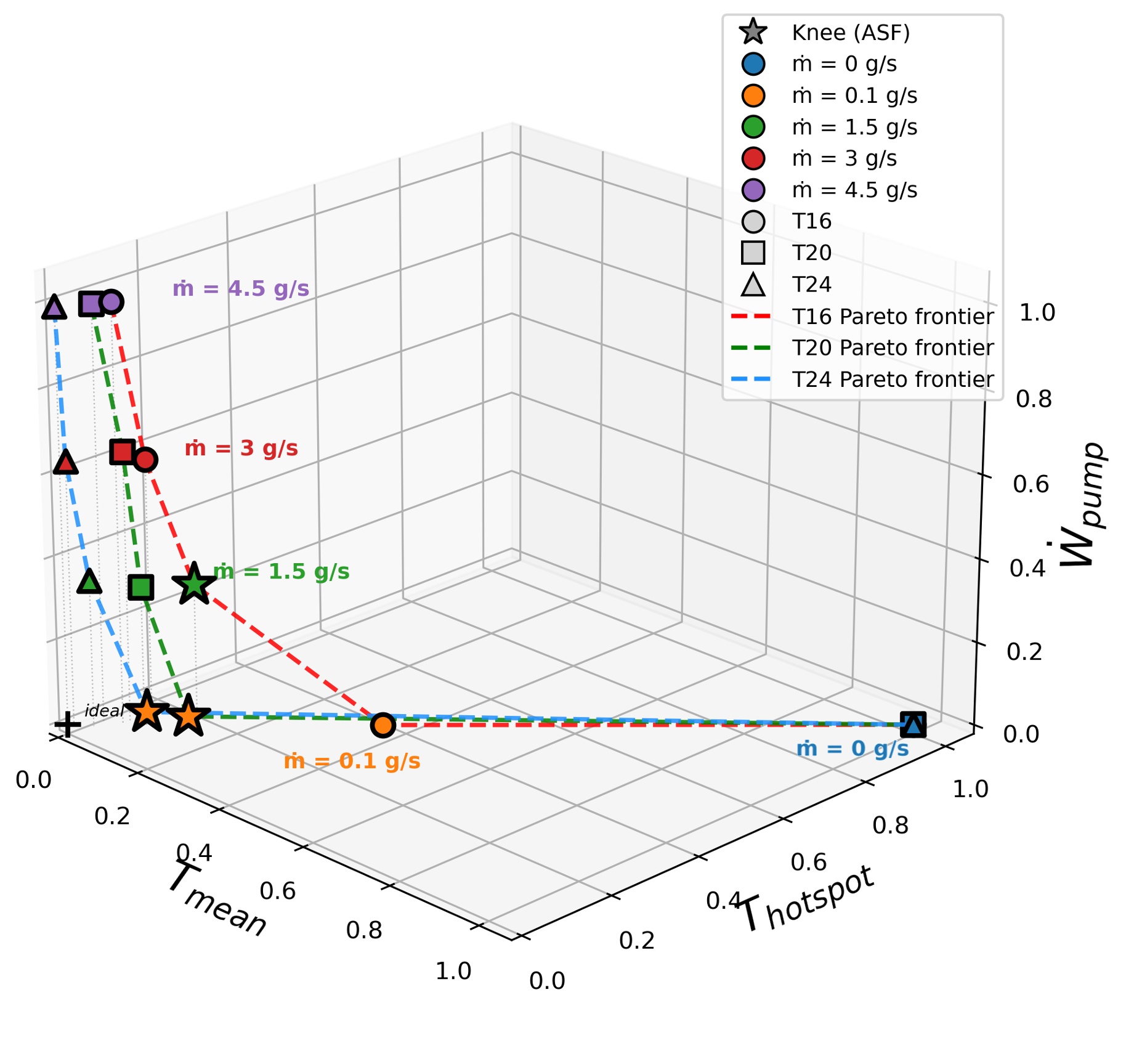}
    \caption{Experimental operating-condition performance map for the manufactured pouch-cell cold plate. Marker shape identifies the nominal inlet-temperature condition (T16, T20, or T24), and colour identifies the nominal coolant mass-flow setting. The axes show globally normalized, baseline-referenced peak-event mean cell temperature, maximum cell temperature, and measured pumping power. Dashed lines connect the non-dominated operating points within each thermal condition; stars identify ASF-selected compromise points. The zero-flow reference is a single experiment displayed in all three groups, not a separate zero-flow measurement at each nominal inlet temperature. Dimensional temperatures and measured inlet conditions are reported in Table~\ref{tab:experimental_absolute}.}
    \label{fig:experimental_pareto_pouch}
\end{figure}

Figure~\ref{fig:experimental_pareto_pouch} maps the measured operating response of the manufactured cold plate. Unlike the simulated geometry Pareto front in Fig.~\ref{fig:pareto3D}, every point in this figure refers to the same machined channel network. The varied input is the nominal coolant mass-flow setting, $\dot m=0$, $0.1$, $1.5$, $3.0$, or $4.5\,\mathrm{g\,s^{-1}}$, under the nominal inlet-temperature conditions T16, T20, and T24. The marker shapes distinguish these conditions and the colours distinguish flow settings. The two thermal coordinates represent peak-event mean and maximum measured cell temperatures relative to the analysis baseline; the third coordinate represents the measured pumping power. Each coordinate is scaled across the assembled dataset for display. Consequently, proximity to the origin describes a joint improvement in the three selected objectives within the tested conditions, not an absolute thermal or energetic target. The dashed lines join operating points identified as non-dominated within each nominal thermal condition. The shared zero-flow observation was measured once and serves as a reference for all three groups; it is not three separate measurements at T16, T20, and T24.

The map exposes the thermal--hydraulic compromise that arises when the cold-plate geometry is held fixed. Introducing coolant flow moves the system away from the high-temperature zero-flow reference, while increasing circulation changes the measured pumping-power coordinate. For T24, the measured maximum cell-temperature rise relative to inlet falls from $6.77\,^{\circ}\mathrm{C}$ at nominal $0.1\,\mathrm{g\,s^{-1}}$ to $6.18$, $5.87$, and $5.56\,^{\circ}\mathrm{C}$ at $1.5$, $3.0$, and $4.5\,\mathrm{g\,s^{-1}}$, respectively. Raising the setting from $3.0$ to $4.5\,\mathrm{g\,s^{-1}}$ therefore reduces the measured maximum rise by another $0.31\,^{\circ}\mathrm{C}$. The corresponding T20 rise decreases from $7.45$ to $6.49\,^{\circ}\mathrm{C}$ between nominal $0.1$ and $4.5\,\mathrm{g\,s^{-1}}$. The T16 response is less orderly: its maximum rise is $7.62\,^{\circ}\mathrm{C}$ at $0.1\,\mathrm{g\,s^{-1}}$, $7.81\,^{\circ}\mathrm{C}$ at $1.5\,\mathrm{g\,s^{-1}}$, $6.92\,^{\circ}\mathrm{C}$ at $3.0\,\mathrm{g\,s^{-1}}$, and $6.65\,^{\circ}\mathrm{C}$ at $4.5\,\mathrm{g\,s^{-1}}$. These increments describe the measured operating points, not the isolated effect of flow at identical inlet temperature and heat load: both inlet temperature and peak-event timing differed among runs. The dimensional temperatures and inlet temperatures in Table~\ref{tab:experimental_absolute} should therefore be consulted alongside the normalized map.

The stars mark the compromise points returned by the achievement scalarizing function (ASF) applied to the normalized objectives. Under this selection rule, the indicated T16 compromise lies near the nominal $1.5\,\mathrm{g\,s^{-1}}$ setting, whereas those for T20 and T24 lie near $0.1\,\mathrm{g\,s^{-1}}$. A different thermal condition changes the relative position of the points on the performance map and can therefore change the selected compromise. In particular, the low-flow T24 point has a maximum measured inlet-relative rise of $6.77\,^{\circ}\mathrm{C}$; the $3.0$ and $4.5\,\mathrm{g\,s^{-1}}$ settings lower that rise to $5.87$ and $5.56\,^{\circ}\mathrm{C}$, respectively. A controller constrained to keep the maximum rise below $6\,^{\circ}\mathrm{C}$ under a comparable T24 load would select from the latter settings rather than the plotted ASF compromise. The difference between these choices reflects the distinction between a balanced multi-objective score and satisfying an explicit temperature limit.

For supervisory thermal control, the measured points provide a finite operating map indexed by nominal thermal condition and commanded coolant flow, with measured inlet temperature, peak cell temperature, spatial sensor-temperature range, and pumping power retained for each setting. Given a measured inlet temperature and an estimate of the current thermal load, the controller can select the operating region represented by the tests, discard flow settings that exceed the allowable peak temperature or spatial-temperature range, and choose the remaining setting with the lowest measured pumping power. Under an elevated thermal load, it can move to a higher-flow setting if the lower-power settings no longer satisfy the thermal limits. In this way the map converts the experimentally observed temperature--power trade-off into a setpoint-selection rule for the manufactured plate. The choice of limits, rather than the location of the ASF star alone, determines the operating point; the map describes the tested operating envelope rather than a demonstrated closed-loop controller.

\subsubsection{Scope and limitations}
The experiments establish thermal response for one manufactured plate under the stated test conditions. Within-window standard deviations do not substitute for independent repeats; sensor-based spatial interpolation limits hotspot localization; and different measured inlet temperatures complicate flow-only comparisons. The unparameterized cell precludes one-to-one validation of absolute coupled-model temperatures. 

\section{Conclusions}

The evolutionary constructal framework produced an organic cold-plate network that was transferred from coupled thermal--electrochemical design to a machined aluminium plate. The pseudo-three-dimensional model evaluates the battery temperature field together with coolant flow and heat transfer, rather than relying on fluid temperature alone. Under matched simulated loading and inlet conditions, the organic design required approximately half the simulated pumping power of the fine parametric straight-channel reference and reduced the battery-temperature nonuniformity objective by up to 30\%. The separately evaluated straight-channel reconstruction retained approximately 60\% of the thermal improvement of the organic design, but was not fabricated or experimentally tested. The experimental hardware corresponded to the organic network.

The pouch-cell experiments characterized the response of the manufactured organic plate over nominal coolant-flow settings of $0.1$--$4.5\,\mathrm{g\,s^{-1}}$ and nominal inlet-temperature conditions T16, T20, and T24. For T24, the maximum measured cell-temperature rise above the respective inlet temperature decreased from $6.77$ to $5.56\,^{\circ}\mathrm{C}$ between the lowest and highest flow settings. Across the tested conditions, the temperature range between instrumented cell locations remained approximately $2.0$--$2.6\,^{\circ}\mathrm{C}$. The experimental performance map combines the measured mean and maximum cell temperatures with the measured pumping power, making the operating trade-off explicit. Its compromise points depend on the thermal condition, the imposed operating constraints, and the objective-selection rule.

The principal value of the framework is that it separates the representation of the electrochemical heat source from the geometry used to distribute and remove that heat. The electrochemical component can therefore be replaced by another validated battery model, heat-generation field, or experimentally prescribed thermal load, while the conjugate cold-plate and evolutionary-design components remain applicable. Similarly, the in-plane design domain and through-thickness coupling formulation can be adapted to different cell form factors, including pouch, prismatic, cylindrical, and module-level configurations, provided that the relevant geometry, thermal interfaces, flow constraints, and boundary conditions are defined. The method thus provides a transferable route for converting spatially resolved thermal loads into manufacturable cooling architectures, while retaining explicit control over the trade-off between temperature uniformity and coolant-power demand.

\subsection*{Acknowledgements}
The authors would like to acknowledge Dr. Waseem Marzook and Dr. Matheus Leal de Souza for training MIC and CN for experiments in the ESE Lab at Imperial.
\subsection*{Declaration of Competing Interest}
The authors declare that they have no known competing financial interests or personal relationships that could have appeared to influence the work reported in this paper.

% \bibliographystyle{cas-model2-names}
% \bibliographystyle{model1-num-names-withURL}
\bibliographystyle{unsrt}
\bibliography{cas-refs}

@article{Alexandersen2020,
   AUTHOR = {Alexandersen, J. and Andreasen, C. S.},
   TITLE = {A Review of Topology Optimisation for Fluid-Based Problems},
   JOURNAL = {Fluids},
   VOLUME = {5},
   YEAR = {2020},
   NUMBER = {1},
   ARTICLE-NUMBER = {29},
   URL = {https://www.mdpi.com/2311-5521/5/1/29},
   ISSN = {2311-5521},
   DOI = {10.3390/fluids5010029}
}

@article{He2024,
   author = {He, L.},
   title = {Spectral Heat Transfer Coefficient for Thermal Design Analysis—Part 1: Augmenting the Cooling Law for Non-Isothermal Wall},
   journal = {Journal of Turbomachinery},
   volume = {147},
   number = {6},
   pages = {061001},
   year = {2024},
   month = {11},
   issn = {0889-504X},
   doi = {10.1115/1.4066671},
   url = {https://doi.org/10.1115/1.4066671},
}

@article{Mo2021,
   author = {X. Mo and H. Zhi and Y. Xiao and H. Hua and L. He},
   doi = {10.1016/J.IJHEATMASSTRANSFER.2021.121612},
   issn = {0017-9310},
   journal = {International Journal of Heat and Mass Transfer},
   month = {10},
   pages = {121612},
   publisher = {Pergamon},
   title = {Topology optimization of cooling plates for battery thermal management},
   volume = {178},
   year = {2021},
}

@inbook{he2023a,
   edition = {1},
   pages = {41-87},
   publisher = {Elsevier},
   title = {Conjugate heat transfer: some fundamentals and recent progress},
   author = {He, L.},
   editor = {Abraham, J. P. and Minkowycz, W. J. and Gorman, J. M.},
   year = {2023},
   book_title = {Advances in Heat Transfer},
   series = {Advances in Heat Transfer}
}

@article{Taeyoung2018,
   author = {Han, T. and Khalighi, B. and Yen, E. C. and Kaushik, S.},
   title = {Li-Ion Battery Pack Thermal Management: Liquid Versus Air Cooling},
   journal = {Journal of Thermal Science and Engineering Applications},
   volume = {11},
   number = {2},
   pages = {021009},
   year = {2018},
   month = {11},
   issn = {1948-5085},
   doi = {10.1115/1.4041595},
   url = {https://doi.org/10.1115/1.4041595},
}

@article{Gungor2024,
   title = {PCM-based passive cooling solution for Li-ion battery pack, a theoretical and numerical study},
   journal = {Applied Thermal Engineering},
   volume = {257},
   pages = {124262},
   year = {2024},
   issn = {1359-4311},
   doi = {https://doi.org/10.1016/j.applthermaleng.2024.124262},
   url = {https://www.sciencedirect.com/science/article/pii/S1359431124019306},
   author = {S. Gungor and S. Lorente},
}

@article{Gungor2022,
   title = {Canopy-to-canopy liquid cooling for the thermal management of lithium-ion batteries, a constructal approach},
   journal = {International Journal of Heat and Mass Transfer},
   volume = {182},
   pages = {121918},
   year = {2022},
   issn = {0017-9310},
   doi = {https://doi.org/10.1016/j.ijheatmasstransfer.2021.121918},
   url = {https://www.sciencedirect.com/science/article/pii/S0017931021010231},
   author = {S. Gungor and E. Cetkin and S. Lorente},
}

@article{Bandhauer2011,
   doi = {10.1149/1.3515880},
   url = {https://dx.doi.org/10.1149/1.3515880},
   year = {2011},
   month = {jan},
   publisher = {The Electrochemical Society, Inc.},
   volume = {158},
   number = {3},
   pages = {R1},
   author = {Bandhauer, T. M. and Garimella, S. and Fuller, T. F.},
   title = {A Critical Review of Thermal Issues in Lithium-Ion Batteries},
   journal = {Journal of The Electrochemical Society},
}

@article{Ning2003,
   title = {Capacity fade study of lithium-ion batteries cycled at high discharge rates},
   journal = {Journal of Power Sources},
   volume = {117},
   number = {1},
   pages = {160-169},
   year = {2003},
   issn = {0378-7753},
   doi = {https://doi.org/10.1016/S0378-7753(03)00029-6},
   url = {https://www.sciencedirect.com/science/article/pii/S0378775303000296},
   author = {G. Ning and B. Haran and B. N. Popov},
}

@article{Wu2015,
   title = {Coupled thermal–electrochemical modelling of uneven heat generation in lithium-ion battery packs},
   journal = {Journal of Power Sources},
   volume = {243},
   pages = {544-554},
   year = {2013},
   issn = {0378-7753},
   doi = {https://doi.org/10.1016/j.jpowsour.2013.05.164},
   url = {https://www.sciencedirect.com/science/article/pii/S0378775313009622},
   author = {B. Wu and V. Yufit and M. Marinescu and G. J. Offer and R. F. Martinez-Botas and N. P. Brandon},
}

@article{Ardani2018,
   title = {Combined experimental and numerical evaluation of the differences between convective and conductive thermal control on the performance of a lithium ion cell},
   journal = {Energy},
   volume = {144},
   pages = {81-97},
   year = {2018},
   issn = {0360-5442},
   doi = {https://doi.org/10.1016/j.energy.2017.12.032},
   url = {https://www.sciencedirect.com/science/article/pii/S0360544217320510},
   author = {M. I. Ardani and Y. Patel and A. Siddiq and G. J. Offer and R. F. Martinez-Botas},
}

@article{Lander2021,
   title = {Cost and carbon footprint reduction of electric vehicle lithium-ion batteries through efficient thermal management},
   journal = {Applied Energy},
   volume = {289},
   pages = {116737},
   year = {2021},
   issn = {0306-2619},
   doi = {https://doi.org/10.1016/j.apenergy.2021.116737},
   url = {https://www.sciencedirect.com/science/article/pii/S0306261921002518},
   author = {L. Lander and E. Kallitsis and A. Hales and J. S. Edge and A. Korre and G. Offer},
}

@article{IgnutaBotas2025,
   author = {M. C. Ignuta-Ciuncanu and R. F. Martinez-Botas},
   doi = {10.1016/J.BIOSYSTEMS.2025.105459},
   issn = {0303-2647},
   journal = {BioSystems},
   month = {4},
   pages = {105459},
   publisher = {Elsevier},
   title = {Discrete Svelteness: Evaluating flow structures in generative constructal design},
   year = {2025}
}

@Inbook{Kozen1992,
   author = {Kozen, D. C.},
   title = {Depth-First and Breadth-First Search},
   bookTitle = {The Design and Analysis of Algorithms},
   year = {1992},
   publisher = {Springer New York},
   address = {New York, NY},
   pages = {19--24},
   isbn = {978-1-4612-4400-4},
   doi = {10.1007/978-1-4612-4400-4_4},
   url = {https://doi.org/10.1007/978-1-4612-4400-4_4}
}

@article{Bejan1978,
   author = {A. Bejan},
   doi = {10.1016/0017-9310(78)90064-9},
   issn = {0017-9310},
   issue = {5},
   journal = {International Journal of Heat and Mass Transfer},
   month = {5},
   pages = {655-658},
   publisher = {Pergamon},
   title = {General criterion for rating heat-exchanger performance},
   volume = {21},
   year = {1978},
}

@article{Bejan2023a,
   title = {Perfection is the enemy of evolution},
   journal = {Biosystems},
   volume = {229},
   pages = {104917},
   year = {2023},
   issn = {0303-2647},
   doi = {https://doi.org/10.1016/j.biosystems.2023.104917},
   url = {https://www.sciencedirect.com/science/article/pii/S0303264723000928},
   author = {A. Bejan},
}

@article{ddpm2020,
   author       = {J. Ho and A. Jain and P. Abbeel},
   title        = {Denoising Diffusion Probabilistic Models},
   journal      = {CoRR},
   volume       = {abs/2006.11239},
   year         = {2020},
   url          = {https://arxiv.org/abs/2006.11239},
   eprinttype    = {arXiv},
   eprint       = {2006.11239},
}

@article{Bejan2023b,
   title = {Constructal design evolution versus topology optimization},
   journal = {International Communications in Heat and Mass Transfer},
   volume = {141},
   pages = {106567},
   year = {2023},
   issn = {0735-1933},
   doi = {https://doi.org/10.1016/j.icheatmasstransfer.2022.106567},
   url = {https://www.sciencedirect.com/science/article/pii/S0735193322006893},
   author = {A. Bejan},
}

@article{Wu2024,
   author = {J. Wu and H. Liu and C. Li and C. Li and G. Xie},
   doi = {10.1016/J.APPLTHERMALENG.2024.123922},
   issn = {1359-4311},
   journal = {Applied Thermal Engineering},
   month = {10},
   pages = {123922},
   publisher = {Pergamon},
   title = {Topological optimization and thermal performance of cold plates for lithium-ion battery with non-uniform heat sources},
   volume = {254},
   year = {2024},
}

@article{Ji2023,
   title = {Topology design of cold plates for pouch battery thermal management considering heat distribution characteristics},
   journal = {Applied Thermal Engineering},
   volume = {224},
   pages = {119940},
   year = {2023},
   issn = {1359-4311},
   doi = {https://doi.org/10.1016/j.applthermaleng.2022.119940},
   url = {https://www.sciencedirect.com/science/article/pii/S1359431122018701},
   author = {H. Ji and T. Luo and L. Dai and Z. He and Q. Wang},
}

@article{Sun2023,
   title = {Thermal design of composite cold plates by topology optimization},
   journal = {International Journal of Mechanical Sciences},
   volume = {259},
   pages = {108594},
   year = {2023},
   issn = {0020-7403},
   doi = {https://doi.org/10.1016/j.ijmecsci.2023.108594},
   url = {https://www.sciencedirect.com/science/article/pii/S0020740323004964},
   author = {C. Sun and W. Wang and X. Tian and X. Zeng and S. Qian and Y. Cai and X. Wang},
}

@article{Shen2022,
   author = {S. C. Shen and M. J. Buehler},
   doi = {10.1038/s44172-022-00037-0},
   issn = {2731-3395},
   issue = {1},
   journal = {Communications Engineering},
   month = {11},
   pages = {1-15},
   publisher = {Nature Publishing Group},
   title = {Nature-inspired architected materials using unsupervised deep learning},
   volume = {1},
   year = {2022},
}

@article{Marquis2019AnAD,
   title = {An Asymptotic Derivation of a Single Particle Model with Electrolyte},
   author = {S. G. Marquis and V. Sulzer and R. Timms and C. P. Please and S. J. Chapman},
   journal = {Journal of The Electrochemical Society},
   year = {2019},
   url = {https://api.semanticscholar.org/CorpusID:168170109}
}

@article{DONDELEWSKI2020,
   title = {The role of cell geometry when selecting tab or surface cooling to minimise cell degradation},
   journal = {eTransportation},
   volume = {5},
   pages = {100073},
   year = {2020},
   issn = {2590-1168},
   doi = {https://doi.org/10.1016/j.etran.2020.100073},
   url = {https://www.sciencedirect.com/science/article/pii/S2590116820300308},
   author = {O. Dondelewski and T. {Szemberg O'Connor} and Y. Zhao and I. A. Hunt and A. Holland and A. Hales and G. J. Offer and Y. Patel},
}

@article{MPM,
   title = {Modelling electrode heterogeneity in lithium-ion batteries: unimodal and bimodal particle-size distributions},
   author = {T. L. Kirk and J. Evans and C. P. Please and S. J. Chapman},
   year = {2020},
   eprint = {2006.12208},
   archivePrefix = {arXiv},
   primaryClass = {physics.app-ph},
   url = {https://arxiv.org/abs/2006.12208},
}

@article{Chu2020,
   title = {Parameterization of prismatic lithium–iron–phosphate cells through a streamlined thermal/electrochemical model},
   journal = {Journal of Power Sources},
   volume = {453},
   pages = {227787},
   year = {2020},
   issn = {0378-7753},
   doi = {https://doi.org/10.1016/j.jpowsour.2020.227787},
   url = {https://www.sciencedirect.com/science/article/pii/S0378775320300902},
   author = {H. N. Chu and S. U. Kim and S. K. Rahimian and J. B. Siegel and C. W. Monroe},
}

@article{Newman_1962,
   doi = {10.1149/1.2425269},
   url = {https://dx.doi.org/10.1149/1.2425269},
   year = {1962},
   month = {dec},
   publisher = {The Electrochemical Society, Inc.},
   volume = {109},
   number = {12},
   pages = {1183},
   author = {Newman, J. S. and Tobias, C. W.},
   title = {Theoretical Analysis of Current Distribution in Porous Electrodes},
   journal = {Journal of The Electrochemical Society},
}

@article{DFN,
   doi = {10.1149/1.2221597},
   url = {https://dx.doi.org/10.1149/1.2221597},
   year = {1993},
   month = {jun},
   publisher = {The Electrochemical Society, Inc.},
   volume = {140},
   number = {6},
   pages = {1526},
   author = {Doyle, M. and Fuller, T. F. and Newman, J.},
   title = {Modeling of Galvanostatic Charge and Discharge of the Lithium/Polymer/Insertion Cell},
   journal = {Journal of The Electrochemical Society},
}

@article{SPM,
   doi = {10.1149/2.0341915jes},
   url = {https://dx.doi.org/10.1149/2.0341915jes},
   year = {2019},
   month = {nov},
   publisher = {The Electrochemical Society},
   volume = {166},
   number = {15},
   pages = {A3693},
   author = {Marquis, S. G. and Sulzer, V. and Timms, R. and Please, C. P. and Chapman, S. J.},
   title = {An Asymptotic Derivation of a Single Particle Model with Electrolyte},
   journal = {Journal of The Electrochemical Society},
}

@article{Pybamm,
   author = {V. Sulzer and S. G. Marquis and R. Timms and M. Robinson and S. J. Chapman},
   doi = {10.5334/JORS.309},
   issn = {20499647},
   journal = {Journal of Open Research Software},
   pages = {1-8},
   publisher = {Ubiquity Press},
   title = {Python Battery Mathematical Modelling (PyBaMM)},
   volume = {9},
   year = {2021},
}

@article{Timms2021,
   author = {R. Timms and S. G. Marquis and V. Sulzer and C. P. Please and S. J. Chapman},
   doi = {10.1137/20M1336898},
   issn = {00361399},
   issue = {3},
   journal = {SIAM Journal on Applied Mathematics},
   month = {5},
   pages = {765-788},
   publisher = {Society for Industrial and Applied Mathematics Publications},
   title = {Asymptotic reduction of a lithium-ion pouch cell model},
   volume = {81},
   year = {2021},
}

@inproceedings{kingma2014autoencoding,
   author = {Kingma, D. P. and Welling, M.},
   booktitle = {ICLR},
   title = {Auto-Encoding Variational Bayes.},
   url = {http://arxiv.org/abs/1312.6114},
   year = 2014
}

@book{bejan_entropy_1982,
   author    = {A. Bejan},
   title     = {Entropy Generation through Heat and Fluid Flow},
   publisher = {Wiley},
   address   = {New York},
   year      = {1982}
}

@article{Bejan1996,
   author = {A. Bejan},
   doi = {10.1063/1.362674},
   issn = {0021-8979},
   issue = {3},
   journal = {Journal of Applied Physics},
   month = {2},
   pages = {1191-1218},
   publisher = {AIP Publishing},
   title = {Entropy generation minimization: The new thermodynamics of finite-size devices and finite-time processes},
   volume = {79},
   year = {1996}
}

@article{Chen2022,
   title = {Topology optimization design and numerical analysis on cold plates for lithium-ion battery thermal management},
   journal = {International Journal of Heat and Mass Transfer},
   volume = {183},
   pages = {122087},
   year = {2022},
   issn = {0017-9310},
   doi = {https://doi.org/10.1016/j.ijheatmasstransfer.2021.122087},
   url = {https://www.sciencedirect.com/science/article/pii/S0017931021011935},
   author = {F. Chen and J. Wang and X. Yang},
}

@article{Fromer2022,
   title = {Computer-aided multi-objective optimization in small molecule discovery},
   journal = {Patterns},
   volume = {4},
   number = {2},
   pages = {100678},
   year = {2023},
   issn = {2666-3899},
   doi = {https://doi.org/10.1016/j.patter.2023.100678},
   url = {https://www.sciencedirect.com/science/article/pii/S2666389923000016},
   author = {J. C. Fromer and C. W. Coley},
}

@article{Khatamsaz2023,
   author = {D. Khatamsaz and B. Vela and P. Singh and D. D. Johnson and D. Allaire and R. Arróyave},
   doi = {10.1038/s41524-023-01006-7},
   issn = {2057-3960},
   issue = {1},
   journal = {npj Computational Materials},
   month = {4},
   pages = {1-14},
   publisher = {Nature Publishing Group},
   title = {Bayesian optimization with active learning of design constraints using an entropy-based approach},
   volume = {9},
   year = {2023},
}

@article{Raabe2023,
   author = {D. Raabe and J. R. Mianroodi and J. Neugebauer},
   doi = {10.1038/s43588-023-00412-7},
   issn = {2662-8457},
   issue = {3},
   journal = {Nature Computational Science},
   month = {3},
   pages = {198-209},
   publisher = {Nature Publishing Group},
   title = {Accelerating the design of compositionally complex materials via physics-informed artificial intelligence},
   volume = {3},
   year = {2023},
}

@article{Bejan2002,
   author = {A. Bejan},
   doi = {10.1016/S0017-9310(02)00165-5},
   issn = {0017-9310},
   issue = {23},
   journal = {International Journal of Heat and Mass Transfer},
   month = {11},
   pages = {4607-4620},
   publisher = {Pergamon},
   title = {Dendritic constructal heat exchanger with small-scale crossflows and larger-scales counterflows},
   volume = {45},
   year = {2002},
}

@article{Bejan2014,
   title = {Constructal design of thermoelectric power packages},
   journal = {International Journal of Heat and Mass Transfer},
   volume = {79},
   pages = {291-299},
   year = {2014},
   issn = {0017-9310},
   doi = {https://doi.org/10.1016/j.ijheatmasstransfer.2014.08.008},
   url = {https://www.sciencedirect.com/science/article/pii/S0017931014006954},
   author = {A. Bejan and S. Lorente and D. Kang},
}

@article{scikit-image,
   title = {scikit-image: image processing in {P}ython},
   author = {van der Walt, S. and Schönberger, J. L. and Nunez-Iglesias, J. and Boulogne, F. and Warner, J. D. and Yager, N. and Gouillart, E. and Yu, T. and the scikit-image contributors},
   year = {2014},
   month = {6},
   volume = {2},
   pages = {e453},
   journal = {PeerJ},
   issn = {2167-8359},
   url = {https://doi.org/10.7717/peerj.453},
   doi = {10.7717/peerj.453}
}

@article{Bejan1997,
   author = {A. Bejan},
   doi = {10.1016/0017-9310(96)00175-5},
   issn = {0017-9310},
   issue = {4},
   journal = {International Journal of Heat and Mass Transfer},
   month = {3},
   pages = {799-816},
   publisher = {Pergamon},
   title = {Constructal-theory network of conducting paths for cooling a heat generating volume},
   volume = {40},
   year = {1997},
}

\end{document}